\UseRawInputEncoding
\pdfoutput=1
\documentclass[letterpaper]{article}
\usepackage[letterpaper, margin=1in]{geometry}

\usepackage{pdfpages}

\usepackage{graphicx} 
\usepackage{amsmath, amssymb} 
\usepackage{geometry} 
\usepackage{subcaption}

\usepackage[font={small},labelfont={bf}]{caption}
 \usepackage[justification=centering]{caption}
 \usepackage{float}

\usepackage{cite}
\usepackage[pagebackref,breaklinks,colorlinks]{hyperref}

\usepackage{tabularx}
\usepackage[inline]{enumitem}
\usepackage{stmaryrd}
\usepackage[squaren]{SIunits}
\usepackage{booktabs}

\title{Optimized Sampling for Non-Line-of-Sight Imaging Using Modified Fast Fourier Transforms}
\author{
    Talha Sultan\\
    University of Wisconsin--Madison, USA\\
    \texttt{tsultan@wisc.edu}
    \and
    Alex Bocchieri\\
    University of Wisconsin--Madison, USA\\
    \texttt{abocchieri@wisc.edu}
    \and
    Chaoying Gu\\
    University of California, Berkeley, USA\\
    \texttt{chaoying\_gu@berkeley.edu}
    \and
    Xiaochun Liu\\
    University of Wisconsin--Madison, USA\\
    \texttt{xliu669@wisc.edu}
    \and
    Pavel Polynkin\\
    University of Arizona, USA\\
    \texttt{ppolynkin@optics.arizona.edu}
    \and
    Andreas Velten\\
    University of Wisconsin--Madison, USA\\
    \texttt{velten@wisc.edu}
}
\date{}

\begin{document}
\maketitle
\begin{abstract}
Non-line-of-Sight (NLOS) imaging systems collect light at a diffuse relay surface and input this measurement into computational algorithms that output a 3D volumetric reconstruction. These algorithms utilize the Fast Fourier Transform (FFT) to accelerate the reconstruction process but require both input and output to be sampled spatially with uniform grids. However, the geometry of NLOS imaging inherently results in non-uniform sampling on the relay surface when using multi-pixel detector arrays, even though such arrays significantly reduce acquisition times. Furthermore, using these arrays increases the data rate required for sensor readout, posing challenges for real-world deployment. In this work, we utilize the phasor field framework to demonstrate that existing NLOS imaging setups typically oversample the relay surface spatially, explaining why the measurement can be compressed without significantly sacrificing reconstruction quality. This enables us to utilize the Non-Uniform Fast Fourier Transform (NUFFT) to reconstruct from sparse measurements acquired from irregularly sampled relay surfaces of arbitrary shapes. Furthermore, we utilize the NUFFT to reconstruct at arbitrary locations in the hidden volume, ensuring flexible sampling schemes for both the input and output. Finally, we utilize the Scaled Fast Fourier Transform (SFFT) to reconstruct larger volumes without increasing the number of samples stored in memory. All algorithms introduced in this paper preserve the computational complexity of FFT-based methods, ensuring scalability for practical NLOS imaging applications.
\end{abstract}


\begin{figure*}
    \centering
    \captionsetup{skip=3pt}
    \includegraphics[width=1\textwidth]{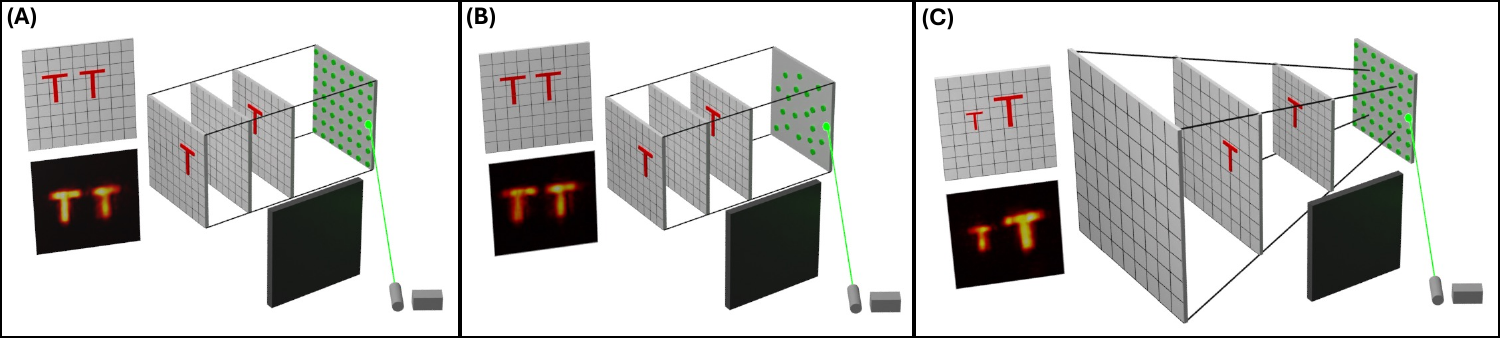}
    \caption{\textbf{(A)} An active imaging system acquires the NLOS measurement using a uniform grid on the relay surface. The standard RSD algorithm reconstructs the hidden scene plane by plane with low computational complexity but constrains both the input and output grids to be uniformly spaced Cartesian grids. We take a max filter along depth to display a 2D image of the 3D reconstruction \textbf{(B)} We develop the NURSD algorithm, which is able to generate 3D reconstructions of the hidden scene acquired with arbitrary input sampling schemes on the relay surface. \textbf{(C)} We develop the SRSD algorithm, which reconstructs larger scenes using the same of samples in memory by increasing the voxel size as a function of depth.}
    \label{fig:Teaser}
\end{figure*}

\section{Introduction}
Non-line-of-sight imaging (NLOS) tackles the challenging problem of recovering the shape and reflectance of objects hidden from the direct line-of-sight (LOS) of an observer (Fig. \ref{fig:Teaser}). Among the various approaches that have been proposed, \cite{faccio_non-line--sight_2020}, Time-of-Flight (ToF) based NLOS imaging is one modality that has demonstrated 3D reconstructions of complex room-sized hidden scenes \cite{liu_non-line--sight_2019, liu_phasor_2020} at 5 Hz \cite{nam_low-latency_2021}. These active imaging systems employ a pulsed laser and time of flight sensors to illuminate and detect multiple spatial locations, respectively, on an intermediary, diffuse relay surface in the LOS of both the system and the hidden scene. At each illumination location, photons scatter off the surface, interact with the hidden objects, and reflect back to the relay surface, where a time-resolved measurement is captured at each detection location (Fig. \ref{fig:Teaser}). Emerging multi-pixel Single Photon Avalanche Diode (SPAD) arrays \cite{renna_fast-gated_2020, riccardo_fast-gated_2022} have been shown to reduce the acquisition times \cite{pei_dynamic_2021, nam_low-latency_2021} by sensing these "three-bounce" photons at multiple detection locations in parallel. 

Computational algorithms take this measurement as an input, and output a 3D reconstruction of the hidden scene sampled at discrete 3D voxels. The fastest algorithms in the literature employ the Fast Fourier Transform (FFT) \cite{otoole_confocal_2018, lindell_wave-based_2019, liu_phasor_2020} to speed up reconstruction. Using the FFT reduces the computational complexity but constrains the sampling scheme for both the input (measurement grid) and the output (voxel grid) to uniformly spaced Cartesian grids (Fig. \ref{fig:Teaser}A). However, capturing these uniform grids is not feasible with multi-pixel arrays, where the input sampling scheme is constrained by the imaging geometry and focusing optics. The output is also constrained to a regular Cartesian grid, which is suboptimal since the lateral imaging resolution degrades linearly with depth \cite{otoole_confocal_2018, dove_paraxial_2019}. Therefore, the reconstruction volume is oversampled at larger depths and generated using an unnecessary amount of samples.

In this paper, we employ extensions of the FFT (Fig. \ref{fig:MFFT_Summary}) to reconstruct the hidden scene with optimized sampling schemes for both the input and the output grids while maintaining the computational complexity of the fastest algorithms. We use the Non-Uniform Fast Fourier Transform (NUFFT) to reconstruct the hidden scene when the measurement is captured with a non-uniformly spaced input grid (Fig. \ref{fig:Teaser}B). Additionally, we utilize the Scaled Fourier Transform (Fig. \ref{fig:Teaser}C) to generate an output grid that accounts for resolution loss by increasing the lateral voxel size linearly with depth. This has the added benefit of generating reconstructions that match human perspective, since objects further away from the relay surface appear smaller (Fig. \ref{fig:Teaser}C).

\subsection{Contributions}
In this paper, we make the following contributions: 
\begin{itemize}
    \item We develop efficient algorithms that reconstruct on datasets acquired with sparse spatial grids with non-uniform spacing on planar and non-planar relay surfaces, reducing strict calibration requirements. 
    \item  We fuse the Scaled Fourier Transform \cite{bailey_fractional_1991} with the standard RSD algorithm \cite{liu_phasor_2020} to develop the SRSD algorithm, which generates reconstruction volumes where the voxel size increases linearly with depth. This enables the processing of larger hidden volumes using less memory and computation relative to existing approaches. Additionally, the increasing voxel size optimally accounts for resolution loss and generates intuitive reconstructions that mirror human perspective.
    \item We utilize the phasor field framework \cite{reza_phasor_2019, reza_phasor_2019-1, teichman_phasor_2019} to mathematically demonstrate that most NLOS imaging setups currently oversample spatially on the relay surface when reconstructing with ToF-based reconstruction algorithms, explaining why many existing approaches are successful in generating high quality reconstructions after discarding many samples in the measurement. This intuition enables us to reconstruct, using the standard RSD algorithm, with roughly 100x less data with minimal change to the reconstruction quality.
\end{itemize}

All algorithms presented in this work preserve the computational complexity of the fastest state-of-the-art algorithms. The remainder of this paper is structured as follows: Section \ref{sec:related-work} reviews related work, highlighting the limitations of existing NLOS imaging methods. Section 3 presents the theoretical framework of our approach. Section \ref{sec:SFFT} - \ref{sec:NUFFT} details our proposed algorithms, while Section \ref{sec:Results} - \ref{sec:NURSD_Results} discusses experimental results validating our methods. Finally, Sections \ref{sec:discussion} - \ref{sec:conclusion} concludes with insights and potential avenues for future research.
\section{Related work}
\label{sec:related-work}

\paragraph{Sampling Schemes and Reconstruction Algorithms} Mathematical operations within the reconstruction algorithms determine the input and output sampling schemes. Algorithms like the filtered backprojection \cite{velten_recovering_2012, arellano_fast_2017, ahn_convolutional_2019} process the measurement and reconstruct sample by sample, allowing for flexible sampling schemes for both the input and output. This flexibility can be incorporated into optimization-based algorithms that iteratively refine where to sample and/or reconstruct \cite{heide_diffuse_2014, manna_error_2018, tsai_beyond_2019, pediredla_snlos_2019, heide_non-line--sight_2019, iseringhausen_non-line--sight_2020, liu_non-line--sight_2021, liu_non-line--sight_2023}.

\begin{figure}
    \centering
    \captionsetup{skip=3pt}  \includegraphics[width=0.7\columnwidth]{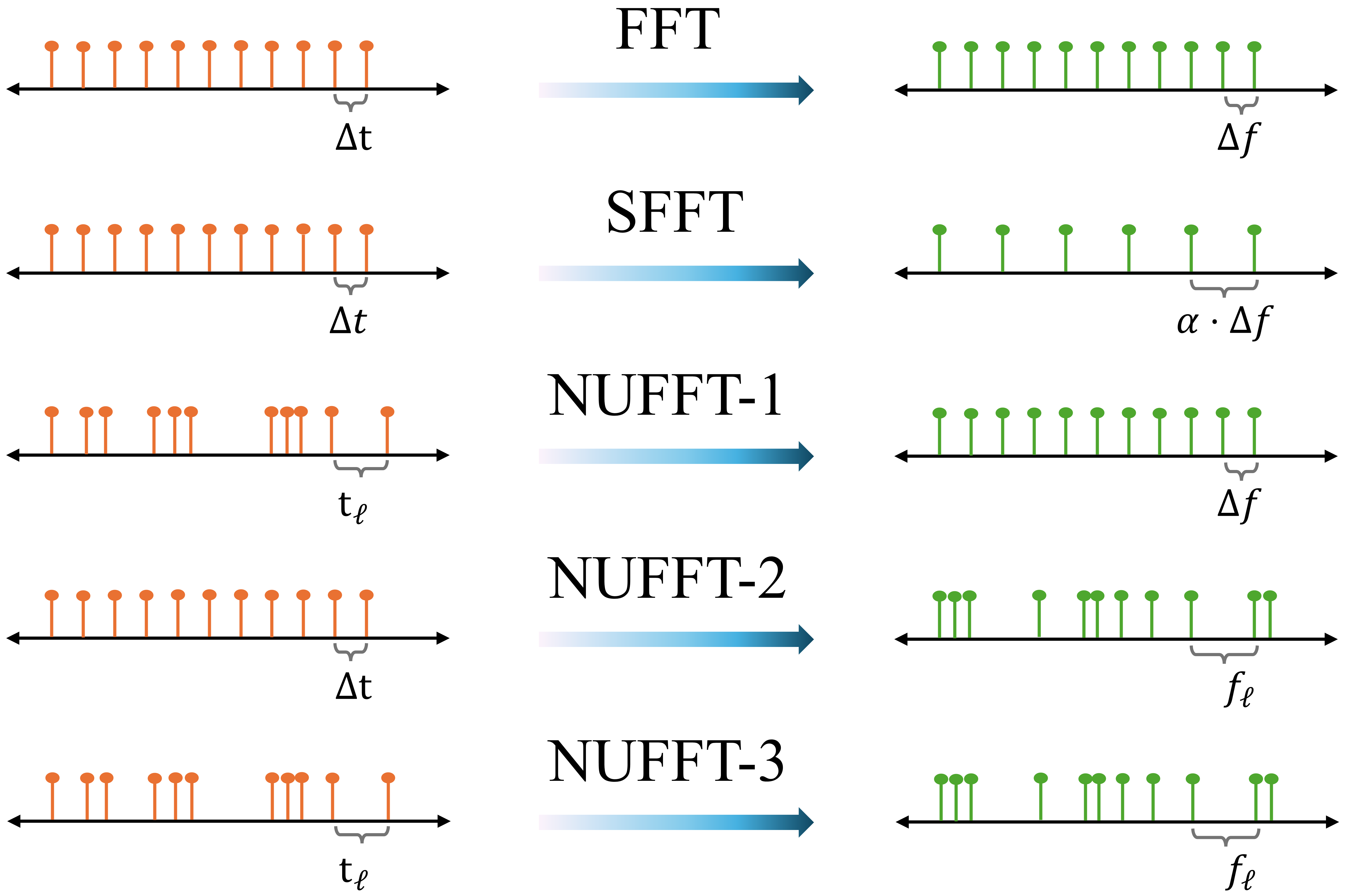}
    \caption{These Modified Fast Fourier Transforms can convert between time $t$ and frequency $f$ with flexible sampling schemes without increasing computational complexity. $\Delta$ is used for uniform spacing, while non-uniform samples are indexed by subscript $\ell$.}
    \label{fig:MFFT_Summary}
\end{figure}

Although this sample-by-sample processing offers flexibility, it incurs significant computational cost. On the other hand, Fourier-based methods formulate the reconstruction operator as a convolution, and then employ the FFT to reduce the computational complexity. For the confocal acquisition scheme, where the laser and detector are focused at the same point before scanning the relay surface, the reconstruction for perfectly diffuse hidden objects can be performed as an FFT-based deconvolution \cite{otoole_confocal_2018, young_non-line--sight_2020, vedaldi_efficient_2020, isogawa_optical_2020}. Wave-based methods, such as the f-k migration algorithm, generalize the reconstruction to hidden objects with arbitrary albedos while using 3D FFTs to maintain the same computational complexity \cite{lindell_wave-based_2019}. The Rayleigh Sommerfeld (RSD) algorithm generalizes the reconstruction to datasets acquired using the full, non-confocal measurement space -  where the illumination position and the detection position are scanned independently on the relay surface - while employing 2D FFTs to maintain the same complexity \cite{liu_phasor_2020}. However, all of these FFT-based methods constrain the input and output sampling schemes to a regular Cartesian grid. This work introduces modified FFTs that enable arbitrary sampling schemes for both the relay surface and the voxel grid (see Fig. \ref{fig:MFFT_Summary}). While we modify the RSD algorithm, we note that these modified FFTs can be incorporated into any FFT-based reconstruction algorithm. 

\paragraph{Phasor Field Framework} This framework allows us to interpret a ToF-based NLOS imaging system as an active, coherent LOS imaging system at the relay surface. Light transmitted by or received at the relay surface can be described as a virtual wavefront. The forward propagation of this wavefront is accurately modeled by diffraction integrals from wave optics \cite{reza_phasor_2019-1, teichman_phasor_2019, dove_paraxial_2019, dove_paraxial_2020, dove_nonparaxial_2020, dove_speckled_2020, reza_phasor_2019, sultan_towards_2024}. These integrals can be used to backpropagate this virtual wavefront into the hidden scene to generate reconstructions of the hidden scene - this forms the basis of the RSD reconstruction algorithm \cite{liu_non-line--sight_2019, liu_phasor_2020}. The phasor field formalism enables us to apply well established principles and techniques from LOS imaging systems, such as cameras, ultrasound, and phased arrays, to understand and enhance NLOS imaging systems. For example, this formalism enables us to establish resolution limits for NLOS reconstructions \cite{dove_paraxial_2019, liu_role_2020}, sampling requirements for NLOS acquisition \cite{liu_role_2020}, provide insights \cite{royo_virtual_2023} into the \textit{missing cone} problem \cite{liu_analysis_2019}, extract complex light transport within the hidden scene \cite{marco_virtual_2021}, and explain formation of virtual mirror images in NLOS reconstructions \cite{royo_virtual_2023}. 

\paragraph{Flexible Input Sampling Schemes} Sample-by-sample processing has been used to reconstruct NLOS measurements acquired on non-planar relay surfaces \cite{manna_non-line--sight-imaging_2020, liu_non-line--sight_2023}, but these algorithms incur significant computational cost. The 3D RSD algorithm employs 3D FFTs to reduce the computational cost to match other FFT based algorithms, and has demonstrated high quality reconstructions of datasets acquired on non-planar relay surfaces \cite{gu_fast_2023}. However, the 3D RSD has two limitations -  first, it is memory inefficient, especially if used for datasets acquired with non-uniform spacing on a \textit{planar} relay surface. Second, it is an approximate solution tends to perform effectively in real-world scenarios. In this work, we employ the NUFFT to develop memory efficient reconstruction algorithms for datasets acquired with non-uniform sampling on both planar and non-planar relay surfaces. We also explain \textit{why} the approximation works well in practice. 

\paragraph{Multi-Pixel Detection Arrays} A uniform input sampling scheme can be generated for the input to the reconstruction algorithm by sequentially scanning the relay surface with a single-pixel detector and/or an illumination source. However, this sequential scanning of a uniform grid increases both the acquisition time and calibration requirements. Recent work demonstrates that SPAD arrays reduce acquisition times since increasing the number of pixels allows for parallel photon acquisition \cite{pei_dynamic_2021, nam_low-latency_2021}. However, detecting a uniformly spaced grid on the relay surface with these arrays is not possible due to the imaging geometry and focusing optics. In this work, we show that the NUFFT can be used to implement efficient reconstruction algorithms for datasets collected with these multi-pixel arrays. Moreover, we show that the reconstruction quality matches that of slower, filtered backprojection algorithm which is considered the state of the art. 

\paragraph{NLOS Compression} An alternate solution to solve the problem of long acquisition times is to undersample the relay surface. Then, there exist two classes of algorithms to tackle this problem. The first approach involves inferring a dense transient measurement from the sparse measurement, usually with deep learning approaches \cite{wang_non-line--sight_2023,li_deep_2023, cho_learning_2024} that require training ahead of time. The second approach aims to reconstruct the hidden scene directly using iterative optimization algorithms \cite{ye_compressed_2021, liu_non-line--sight_2023} that increase the computational complexity. However, the question of \textit{why} the measurement is compressible along the spatial dimension has not yet been addressed. 

In this work, we utilize the phasor field framework to mathematically demonstrate that the NLOS can be undersampled along the relay wall, provided we are not limited by Poisson Noise. Furthermore, we devise a simple algorithm that interpolates undersampled data and generates 3D reconstructions without significant loss in reconstruction quality. The computational complexity then depends on the interpolation technique but we show that even using nearest neighbor interpolation yields sufficiently good results. Our goal is not to outperform existing approaches but to provide a justification for why the first class of methods tend to work fairly well, and motivate the use of the Non-Uniform Fourier Transform for NLOS imaging with SPAD arrays.

\paragraph{Computational Holography} Holography is another area where a coherent wavefront can be extracted from a computed hologram or from a hologram recorded on a digital sensor \cite{10.1093/jmicro/dfy007, 10.1145/3378444}. Diffraction algorithms generate a 3D reconstruction by backpropagating this wavefront into a reconstruction volume. Modified Fourier Transforms discussed in this paper have been previously fused with diffraction algorithms in the context of holography \cite{muffoletto_shifted_2007, shimobaba_nonuniform_2013}. However, these modified FFTs are implemented for the Fresnel Diffraction Integral, which is an approximation to the more accurate RSD integral used in this work\cite{goodman_introduction_2005}.

\section{Background: Phasor-field Imaging}
\label{sec:background}

Our methods are based on the phasor field formalism. This framework transforms the NLOS measurement into a complex wavefront collected by an LOS coherent imaging system. Diffraction integrals from wave optics can be utilized to backpropagate the wavefront into the hidden volume, generating 3D reconstructions of the hidden objects. In the next section, we describe the fast RSD algorithm \cite{liu_non-line--sight_2021} that operates on the temporal frequency domain of the measurement. Since our contributions build upon and extend this algorithm, it is essential to review it in detail. 

\subsection{Standard RSD Algorithm}
The RSD algorithm is able to generate reconstructions of the hidden scene from both confocal and non-confocal measurements and can be accelerated using the FFT \cite{liu_non-line--sight_2021, nam_low-latency_2021}. The main idea of the RSD is that it encodes the temporal shifts in the measurement as phasors in the Fourier domain, so time shifts in the reconstruction algorithm can be computed using complex addition. Furthermore, Fourier-based convolution can be used to describe plane-to-plane diffraction and accelerate the calculation using the Fast Fourier Transform (FFT). Consequently, the RSD algorithm maintains accuracy only when both the planar relay wall and each reconstruction plane are sampled uniformly. 

The transient measurement, \( H(\vec{x}_p, \vec{x}_c, t) \), is recorded for various laser positions \( \vec{x}_p \) and camera positions \( \vec{x}_c \) on the relay surface at times \( t \). Since the light transport for each spatial location in our imaging system is linear and time-invariant, the transient measurement can be formulated as the impulse response of the hidden scene. Thus, the virtual response, $\mathcal{P}(\vec{x}_p, \vec{x}_c, t)$, of the hidden scene to any virtual illumination can be computed by convolving the impulse response with the corresponding illumination function, $\mathcal{P}(\vec{x}_p, t)$:

\begin{equation}
\begin{aligned}
\mathcal{P} (\vec{x}_p, \vec{x}_c, t) = \mathcal{P} \left(\vec{x}_p, t\right) * H(\vec{x}_p, \vec{x}_c, t)
\label{eq:PF_Response}
\end{aligned}
\end{equation}

where each element in the temporal frequency, $\omega$, domain of the virtual response, $\mathcal{P_F} (\vec{x}_p, \vec{x}_c, \omega)$, is a phasor. This is a complex number that encodes the temporal phase accumulated by a monochromatic phasor field wavefront as it propagates from an illumination position, $\vec{x}_p$, into the hidden scene and is reflected back to a detection position $\vec{x}_c$. We then utilize the RSD integral to backpropagate each wavefront into the hidden scene to generate 3D reconstructions, $I\left(\vec{x}_{{v}}\right)$, at voxel positions $\vec{x}_{v}$:  

\begin{equation}
\begin{aligned}
    \label{eq:PF_RSD_Defn}
   I\left(\vec{x}_{{v}}\right) 
    &= \int \underbrace{e^{-j \frac{\omega}{c}\left|\vec{x}_{{p}}-\vec{x}_{{v}}\right|}}_{\mathcal{P_F}(\vec{x}_p)}  
    \int \underbrace{\left[\int  \mathcal{P_F} \left(\vec{x}_p, \vec{x}_c, \omega\right)
     e^{-j \frac{\omega}{c}\left|\vec{x}_{{v}}-\vec{x}_{{c}} \right|}  d \vec{x}_{{c}} 
    \right]}_{R_{\vec{x}_v} \left(\mathcal{P_F} \left(\vec{x}_c, \omega\right)\right)} d \omega  d \vec{x}_{{p}}   \\
\end{aligned}
\end{equation}

where $\mathcal{P_F}(\vec{x}_p)$ is an illumination position dependent phase mask, and $R_{x_v} (\mathcal{P_F} \left(\vec{x}_c, \omega\right))$ is a detection position dependent phase mask. The computation of $R{\vec{x}_v}$ can be accelerated by using FFT based convolution, provided all the detection points $\vec{x}_c$ are uniformly sampled on a planar surface: 

\begin{equation}
\begin{aligned}
R_{\vec{x}_{{v}}}  \left(\mathcal{P}_{\mathcal{F}}\left({\vec{x}}_{{c}}, \omega\right)\right) 
       &=   \mathcal{P}_{\mathcal{F}}\left(
        x_c, y_c, 0, \omega \right) \underset{\text{2D}}{*} G\left(x_c, y_c, z_v, \omega\right) \\  
\end{aligned}
\label{eq:RSD_conv}
\end{equation}

where $ G\left(x_c, y_c, z_v, \omega \right) = {e^{i\frac{\omega}{c} \sqrt{{x_c}^2 + y_c^2 + z_v^2}}}/{\sqrt{x_c^2 + y_c^2 + z_v^2}} $ is the propagation kernel, and is invariant for each $z_v$. Therefore, all voxel $\vec{x}_v$ can be reconstructed simultaneously for a given plane and so the 3D volume can be generated plane by plane. 

Helmholtz reciprocity can be used to interchange illumination and detection positions \cite{veach_robust_1997, sen_dual_2005} so the SPAD pixel becomes a virtual phasor field illumination source and Eq \ref{eq:RSD_conv} can be used on the laser positions, provided the laser grid is scanned with uniform spacing on the relay wall \cite{liu_phasor_2020}. 

For a uniform sampling grid, each discrete location can be formulated as an integer multiple of the sampling rates in each dimension i.e. $(m \Delta x, n\Delta y)$, where $(\Delta x, \Delta y)$, are the sampling rates and  $(m , n)$ are integers s.t. $m \in ( -M/2, +M/2 )$ ,  $n \in ( -N/2, +N/2 )$. We can rewrite our plane to plane convolution as: 

\begin{equation}
 \begin{split} 
\mathcal{P}_{\mathcal{F}} \left( m\Delta x, n\Delta y, z_v, \omega \right) 
&= \\
\operatorname{IFFT}\Bigl\{ \operatorname{FFT}\left\{\mathcal{P}_{\mathcal{F}} \left( m\Delta x, n\Delta y, 0, \omega \right) \right\}  & \cdot  \operatorname{FFT}\left\{G\left(m\Delta x, n\Delta y, z_v, \omega\right) \right\}\Bigr\}
\label{eq:Discrete_RSD_Conv}
\end{split}   
\end{equation}

We report the complexity using the commonly used notation available in the literature \cite{liu_phasor_2020}. The reconstruction volume is sampled at the same rate as the relay surface, though zero padding can be used to increase the range and generate reconstructions over a larger lateral distance. Suppose the side length of the reconstructed cube is $N$. Then, we need to perform 2D FFTs a total of $N$ times for a fixed number of frequency components, giving a computational complexity of $O(N^{3} \log N)$ for a single illumination position $\vec{x}_p$. To differentiate this from the modified algorithms introduced in this paper, we will henceforth refer to this algorithm as the Standard RSD algorithm. The filtered backprojection (FBP) algorithm reconstructs the same sized volume by solving Eq \ref{eq:PF_RSD_Defn} directly with $O(N^5)$  complexity \cite{liu_non-line--sight_2019}.

\subsection{3D RSD}
The 3D RSD is an extension of the Standard RSD algorithm that enables fast reconstruction of NLOS data that is measured on nonplanar relay surfaces with non-uniform sampling schemes \cite{gu_fast_2023}. The 3D RSD algorithm generates reconstruction quality that is comparable to prior state of art while being orders of magnitude faster due to its lower computational complexity. 

This algorithm proceeds in 2 stages. In Stage 1, the virtual response, $\mathcal{P_F} \left( {\vec{x}_{\mathrm{c}}}^\prime, \omega \right)$, which is collected on a non-uniform grid ${\vec{x}_{\mathrm{c}}}^\prime$, is interpolated ($\phi(.)$) to a uniform 3D grid represented by $\vec{x}_c = (x_c, y_c, z_c)$:

\begin{equation}
\begin{aligned}
& \mathcal{P_F} \left(\vec{x}_{\mathrm{c}}, \omega \right)
= \mathcal{\phi} \left( \mathcal{P_F} \left( {\vec{x}_{\mathrm{c}}}^\prime, \omega  \right) \right)
\end{aligned}
\label{eq:3d_rsd_Int}
\end{equation}

followed by propagation to an intermediate plane, $z_0$. This propagation can be written as a 3D convolution and accelerated using 3D FFTs:

\begin{equation}
\begin{aligned}
& \int \mathcal{P_F} \left(\vec{x}_{{c}}, \omega\right) e^{-j \frac{\omega}{c}\left|\vec{x}_{{v}}-\vec{x}_{{c}}\right|} d \vec{x}_{{c}} \\
= & \sum_{x_c}\sum_{y_c}\sum_{z_c} \mathcal{P_F}\left(x_c, y_c, z_c, \omega\right) e^{-j \frac{\omega}{c} \sqrt{\left(x_v-x_c\right)^2+\left(y_v-y_c\right)^2+\left(z_v-z_c\right)^2}} 
\\
= & \left. \mathcal{P_F}\left(x_v, y_v, z_v -z_0, \omega\right) \underset{\text{3D}}{*}
G(x_v, y_v, z_v, \omega) \right|_{z_v = z_0}
\end{aligned}
\label{eq:3d_rsd}
\end{equation}

where $G(x_v, y_v, z_v, \omega) = e^{-j \frac{\omega}{c}\sqrt{\left(x_v\right)^2+\left(y_v\right)^2+\left(z_v\right)^2}} $.

In Stage 2, we utilize the plane to plane RSD algorithm to propagate the wavefront at $z_0$ to the remaining planes in the reconstruction volume. This relies on the well known separability of the RSD \cite{born_principles_2019}.

Since both stages can be accelerated using Fourier-based convolution, the algorithm preserves the computational complexity of $O(N^{3} \log N)$, provided that the complexity of the interpolation operator, $\phi(.)$, is bounded.

\subsection{Phasor Field Sampling, Filtering, and Resolution}

While the phasor field illumination function can be any arbitrary function, we set it as a bandpass filter centered around $\omega_c$ in the frequency domain:

\begin{equation}
\begin{aligned}
\mathcal{P_F}\left(\vec{x}_p, \omega \right)&=   \delta \left(\vec{x} - \vec{x}_p \right) e^{-\frac{(\omega - \omega_c)^{2} }{2\sigma^2}} 
\label{eq:PFPacket_Omega_Supp}
\end{aligned}
\end{equation}

where the phasor field central wavelength, \( \lambda_c \), can be extracted from \( \omega_c \) using  \( \lambda_c =  2 \pi c/\omega_c \), $c$ is the speed of light, and filter width is governed by  \( \sigma = c/(5 \lambda_c) \). Provided that the relay wall is sampled at $\lambda_s = \lambda_c/2$ \cite{liu_non-line--sight_2019} to satisfy nyquist criterion, the lateral reconstruction resolution, $\Delta x$, for a reconstructed point source in the hidden scene has been shown to be \cite{reza_phasor_2019, dove_paraxial_2019, dove_paraxial_2020, sultan_towards_2024}:

\begin{equation}
\begin{aligned}  
\Delta x &= \frac{1.22\lambda_c z}{D} 
\label{eq:Focusing_Res}
\end{aligned}
\end{equation}

where $z$ is the depth offset from the relay surface and $D$ is the diameter of the relay wall. The lateral resolution therefore degrades linearly with both $\lambda_c$ and $z$, demonstrating how the standard RSD algorithm oversamples at larger depths. 

\paragraph{Denoising/Filtering} Increasing \( \lambda_c \) "denoises" the reconstruction at the expense of reconstruction quality, since the filter increasingly selects lower frequencies and fewer frequencies as both \( \omega_c \) and \(\sigma\) decrease. In Section \ref{sec:SubSampling}, we show how adjusting $\lambda_c$ can denoise the measurement and enable interpolation of noisy data. This allows us to use fewer measurements than prescribed by the Nyquist criterion.
\section{Methods: Scaled RSD}
\label{sec:SFFT}
We introduce the Scaled RSD as an extension of the original RSD that computes plane-to-plane diffraction. Using an FFT to accelerate the reconstruction means that the spatial sampling rate on the planar relay surface, $x_c$, determines the output sampling rate of each reconstructed plane. Our goal is to computationally change the sampling rate of the output plane without changing the sampling rate at the input plane, since the latter is governed by hardware and maybe difficult to modify physically. Furthermore, this modification should not change the underlying computationally complexity of the reconstruction algorithm.

\subsection{Math of Scaled RSD}
We want to query the phasor field at a scaled grid $(\alpha x_c, \beta y_c)$ instead of the regular grid $(x_c, y_c)$, so we rewrite Eq. \ref{eq:RSD_conv}:

\begin{equation}
\begin{aligned}
\mathcal{P}_{\mathcal{F}}\left(
        \alpha x_c, \beta y_c, z_v, \omega \right) &=   \mathcal{P}_{\mathcal{F}}\left(
        \alpha x_c, \beta y_c, 0, \omega \right) \underset{\text{2D}}{*} G\left(\alpha x_c, \beta y_c, z_v, \omega\right) 
\end{aligned}
\label{eq:SRSD_conv}
\end{equation}

For the convolutional kernel, we can easily calculate with the scaled grid using its definition:
\begin{equation}
\begin{aligned}
    G\left(\alpha x_c, \beta y_c, z_v, \omega \right) = {e^{i\frac{\omega}{c} \sqrt{{(\alpha x_c)}^2 + (\beta y_c)^2 + z_v^2}}}/{\sqrt{(\alpha x_c)^2 + (\beta y_c)^2 + z_v^2}}
\end{aligned}
\end{equation}

 To evaluate the phasor field on the scaled grid $(\alpha x_c, \beta y_c)$, we utilize the Scaled Fast Fourier Transform (SFFT) - also known as Fractional Fast Fourier Transform in the literature \cite{bailey_fractional_1991, 1162132}. This allows for non-integer scaling of the frequency domain representation of an input wavefront. Computing a 2D SFFT requires the input signal to be uniformly sampled:
 
\begin{equation}
\begin{split}
\mathcal{P}_{\mathcal{F}} \left( x_c,  y_c, z_v, \omega \right) 
 & = \mathcal{P}_{\mathcal{F}} \left( m\Delta x, n\Delta y, 0, \omega \right) \\
& (m , n) \in \mathcal{I}_{M, N}
\end{split}
\end{equation}

where we define 

\begin{equation}
\begin{split}
I_\mathcal{L} := \{ \ell \in \mathcal{Z}: -\mathcal{L}/2 \leq \ell < \mathcal{L}/2  \}
\end{split}
\end{equation}

so $\mathcal{I}_{M, N}$ is a 2D grid. Defining the 2D DFT kernel as:

\begin{equation}
\begin{aligned}
\mathrm{U}\left[m^{\prime}, n^{\prime}\right] 
&\triangleq \sum_{m=-M / 2}^{M / 2-1} \sum_{n=-N / 2}^{N / 2-1} \mathrm{u}[m, n] \cdot  k_M^{- m \cdot m^{\prime}}  k_N^{- n \cdot n^{\prime}} 
\end{aligned}
\label{eq:2D_DFT}
\end{equation}

where 

\begin{equation}
  \begin{split}
 k_\mathcal{L}=\exp \left[-j \frac{2 \pi}{\mathcal{L}} \right]
\end{split}
\end{equation}

In Supplement Section \ref{sec:SFFT_Math}, we derive that the SFFT, i.e. the spectrum of the input phasor field kernel in Eq \ref{eq:Discrete_RSD_Conv} on the scaled grid $(\alpha x_c, \beta y_c)$, can be evaluated using 3 additional FFTs:

\begin{equation}
 \begin{split} 
 \operatorname{FFT} \left\{
\mathcal{P}_{\mathcal{F}} \left( \alpha n\Delta x,\beta  n\Delta y, z_v, \omega \right) \right\}
&=  k_M^{-\frac{\alpha}{2} m^{\prime 2}} \cdot k_N^{-\frac{\beta}{2} n^{\prime 2}}
\\ 
\operatorname{IFFT}\Bigl\{ \operatorname{FFT}\left\{\mathcal{P}_{\mathcal{F}}^{\prime} \left( n\Delta x, n\Delta y, 0, \omega \right) \right\}  &
 \cdot  \operatorname{FFT}\left\{   k_M^{\frac{\alpha}{2} m^2} \cdot k_N^{\frac{\beta}{2} n^2}
 \right\}\Bigr\}
 \label{eq:SFFT}
\end{split}   
\end{equation}

where

\begin{equation}
\mathcal{P}_{\mathcal{F}}^{\prime} \left( n\Delta x, n\Delta y, 0, \omega \right) = \mathcal{P}_{\mathcal{F}} \left( n\Delta x, n\Delta y, 0, \omega \right) \cdot  k_M^{-\frac{\alpha}{2} m^{ 2}} \cdot k_N^{-\frac{\beta}{2} n^{2}}
\end{equation}

Computing three additional FFTs does not increase the computational complexity. In the next few sections, we demonstrate instead that this improves the overall complexity since the Scaled RSD reconstructs a larger volume using the same number of samples. 

\begin{figure}
    \centering
    \captionsetup{skip=3pt}  \includegraphics[width=0.7\columnwidth]{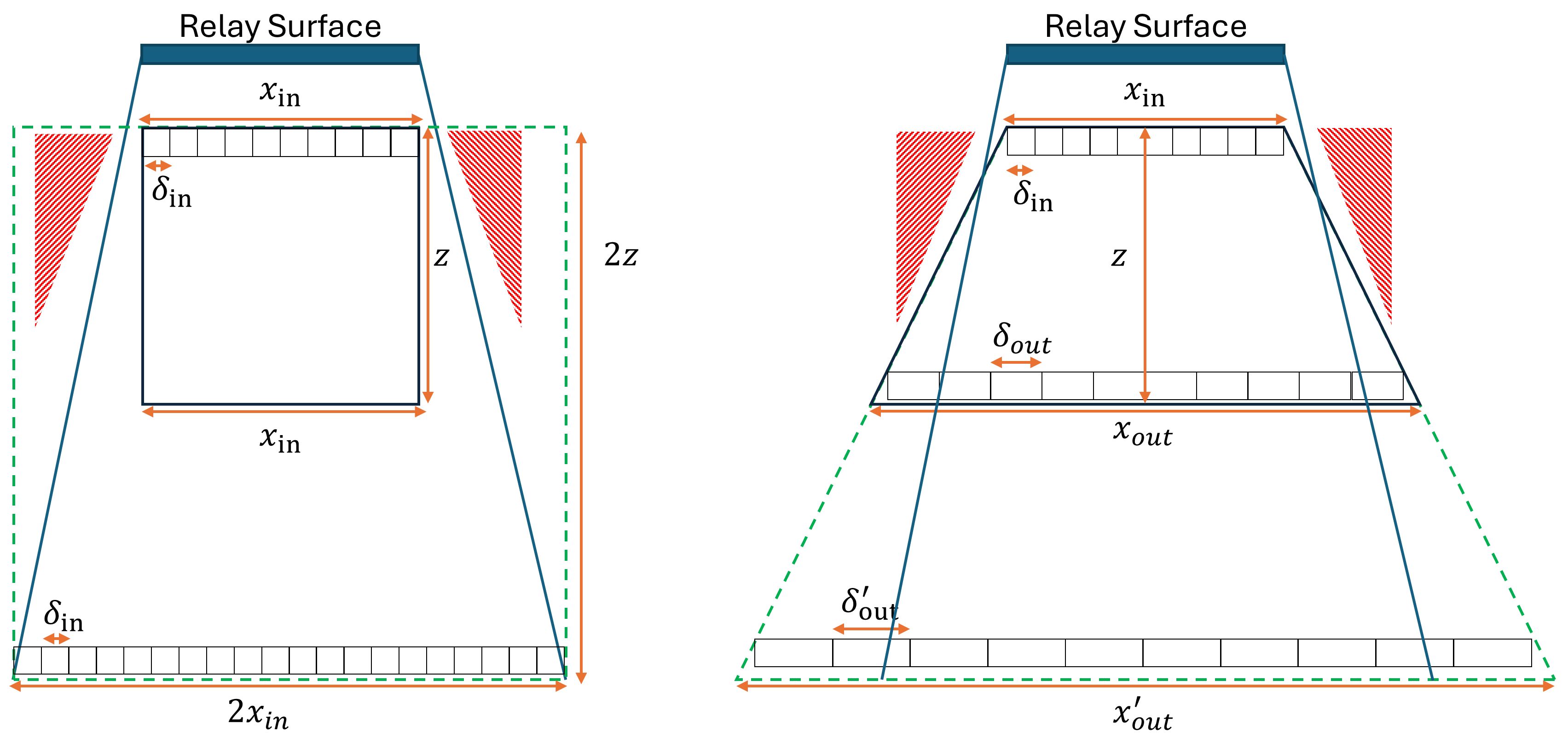}
    \caption{Showing the top view of the reconstruction volume when reconstructing with the RSD algorithm (left) and the SRSD algorithm (right). The dotted green lines demonstrate the increase in volume as we increase $z$, when the lateral field of view (FOV) is fixed. The RSD algorithm requires zero padding and increasing the side length, while the SRSD increases the lateral FOV without increasing the number of pixels.}
    \label{fig:SRSD_Complexity}
\end{figure}

\paragraph{Scaling Factors} The scaling factors can be adjusted to either increase or decrease the voxel size. In this work, we set scaling factors $\alpha, \beta$ between 0 and 1 so that the side lengths, $x_{out}(z)$ and $y_{out}(z)$, of the reconstruction volume, increase linearly with $z$:

\begin{align}
x_{out}(z) &= x_{in} + \frac{z}{\alpha}\ \ , \qquad 0\leq \alpha \leq 1  \label{eq:x_out}\\
y_{out}(z) &= y_{in} + \frac{z}{\beta}\ \ , \qquad 0\leq \beta \leq 1  \\
z_{out} &= z_{in} + z 
\end{align}

where $x_{in}, y_{in}$ indicate the initial lateral side lengths (Fig. \ref{fig:SRSD_Complexity}), and $z_{in}$ and $z_{out}$ are the initial and final depths from the relay surface. As shown in Fig. \ref{fig:Teaser} and Fig. \ref{fig:SRSD_Complexity}, the shape of the reconstruction volume changes from a \textit{cuboid} to a \textit{pyramidal frustum}. 

For simplicity, we assume that $x_{out} = y_{out}$ for any $z$ so $\alpha = \beta$, and the cross-section of this frustum at any $z$ is a square. We derive, in Supplement Section \ref{sec:Volume_Frustum}, that the volume of this specific frustum is defined by: 

\begin{align}
V_F &= z x^2_{in}  + z^2 \frac{x_{in} }{\alpha} + \frac{z^3}{3\alpha^2}
\end{align}

where $z$ is the height of this frustum. As shown in Fig. \ref{fig:SRSD_Complexity}, the output pixel side length, $\delta_{out}(z)$, increases proportionally with $z$ (Fig. \ref{fig:SRSD_Complexity}), and is given by:

\begin{align}
\delta_{out} (z) &= \frac{\delta_{in}}{\alpha (z)} = 
\left(\frac{x_{out}(z)}{ x_{in}}\right)  \delta_{in}
\end{align}

where $\alpha(z)$ is the varying scaling factor that incorporates the $z$ dependence from Eq \ref{eq:x_out}.  

Increasing the output voxel size compensates for the resolution loss while reducing the number of samples needed for reconstruction. This raises a key question: what is the maximum voxel size that can be used for reconstruction without losing information? The lateral resolution from Eq.~\ref{eq:Focusing_Res} provides a basis for deriving upper and lower bounds respectively for $\delta_{out}(z)$ and $\alpha (z)$:

\begin{align}
\delta_{out}(z) = \frac{\delta_{in}}{\alpha (z)} & \leq \frac{\Delta x }{2} = \frac{1.22 \lambda_c z}{2D} \\
\alpha (z)  & \geq  \frac{2 \delta_{in}}{\Delta x} 
\end{align}

\subsection{Complexity Analysis}
The computational complexity reported in the literature assumes the reconstruction volume is a cube. For the standard RSD, this volume is equal to $V_C = z^3$ since $z = x_{in} = y_{in}$. The number of samples used in memory is equal to $Z^3 = (z/\delta)^3$. As shown by the solid black lines in Fig. \ref{fig:SRSD_Complexity}, we note that the SRSD algorithm uses the same number of samples to reconstruct a larger volume and the difference is governed by 

\begin{equation}
\begin{aligned}
\Delta V & = V_F - V_C = \left( \frac{1}{\alpha}+\frac{1}{3\alpha^2} \right) z^3
\label{eq:Vol_Frustum}
\end{aligned}
\end{equation}

when $x_{in} = z$. As $\alpha$ approaches infinity, the difference goes to zero making the two volumes identical. As a concrete example, if $\alpha=0.5,\ z = 4\ m$, then $V_C = 64\ m^3$, $V_F = 277\ m^3$, $\Delta V = 213\ m^3$, and the percentage increase in volume, $\Delta V/V_C$, is $333\%$.

We analyze the computational complexity for a fixed field of view (FOV) as a function of depth, as indicated by the two blue lines in Fig. \ref{fig:SRSD_Complexity}, for a virtual camera placed behind the relay wall. The existing implementation of the standard RSD \cite{liu_phasor_2020} requires zero padding to increase the side length to match the field of view (Fig. \ref{fig:SRSD_Complexity} left), before performing three 2D FFT \textbf{(}$O(Z^2 \log Z)$ steps\textbf{)} to reconstruct each plane. Therefore, the number of output samples/voxels scale by $O(Z^3)$ while the computation scales by $O(Z^3\log Z)$. The SRSD, by contrast, adjusts the output sample size ($\delta_{\text{out}}$ in Fig. \ref{fig:SRSD_Complexity}) at each depth to match the required FOV by computing 3 additional FFTs. This approach adds a constant number of samples per depth, and computes a total of six 2D FFTs for each depth for $N^2 = (x/\delta_{in})^2$ output samples, matching the number of input samples on the relay surface (Fig. 1\ref{fig:SRSD_Complexity}). Therefore, the number of output samples scale by $O(Z)$ while the computational complexity scales by $O(Z N^2\log N)$. 

We note that the standard RSD algorithm can be easily modified  to output a frustum by changing the zero padding as a function of depth. This adjusted implementation can reduce the number of samples used in memory, but is still suboptimal since the pixel size remains fixed and therefore, does not account for resolution loss as $z$ increases. Additionally, the standard RSD can achieve the same complexity as the SRSD by removing the FOV requirement and reconstructing the same size plane at each depth but this means we reconstruct a significantly smaller volume.  

Furthermore, Fig. \ref{fig:SRSD_Complexity} shows that the standard RSD reconstructs a larger lateral area, marked in red, closer to the relay surface which the SRSD ignores. This is inefficient since hidden objects located at these steep angles reflect limited amount of photons back to the relay surface, and can not be reconstructed with the same resolution \cite{liu_analysis_2019, royo_virtual_2023}. Furthermore, this area may be in the line of sight of the imaging system.

\begin{figure*}
    \centering
\captionsetup{skip=3pt}
\includegraphics[width=1\textwidth]{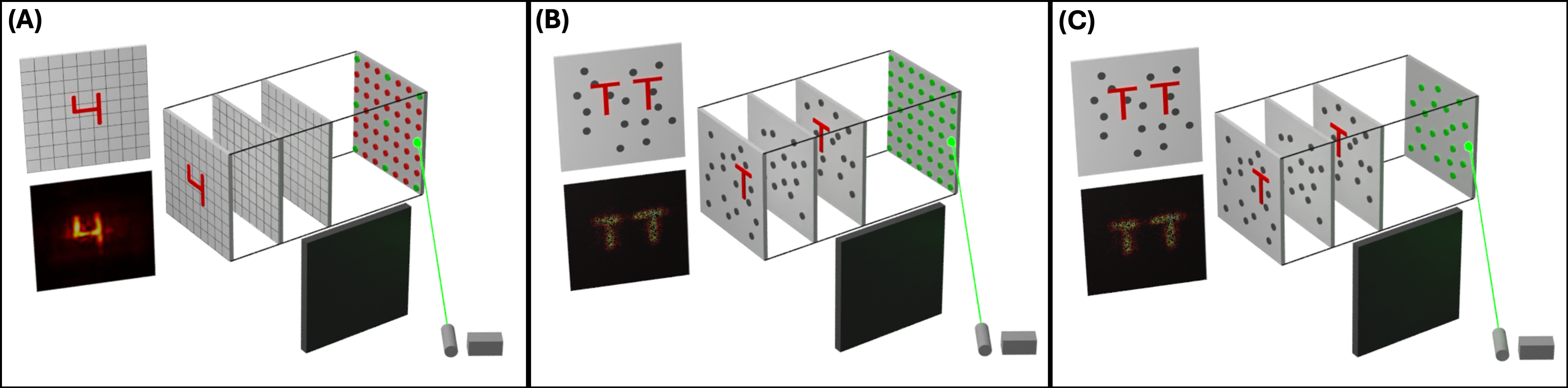}\caption{\textbf{(A)} The relay surface is sparsely sampled at the green locations and interpolated to a dense grid denoted by the red locations. Reconstructing the hidden scene using the Standard RSD algorithm on the interpolated grid generates an output with minimal loss in reconstruction quality. \textbf{(B)} The NURSD-2 algorithm can generate reconstructions where the output i.e. the voxel grid is processed with flexible sampling schemes. \textbf{(C)} The NURSD-2 algorithm can generate reconstructions where both the input and output are discretized with flexible sampling schemes.}
\label{fig:Teaser_2}
\end{figure*}

\section{Methods: Sampling Criteria for NLOS}
\label{sec:NLOS_CS}

To solve the problem of limited pixels and high data rates from increasing the number of SPAD pixels, we exploit redundancy in NLOS measurements. The standard RSD algorithm \cite{liu_phasor_2020} requires spatial sampling at $\lambda_s = \lambda/2$ to satisfy the Nyquist criteria, where $\lambda^*$ is the wavelength corresponding to the maximum frequency in the phasor field kernel.   

Suppose we have a phasor field point source in the hidden scene oscillating at $\omega = 2 \pi c / \lambda_c$. We show in Supplement section \ref{sec:NLOS_CS_Proof} that while the depth axis ($z$) must always be sampled at $\lambda_{sz} = \lambda^*/2$, the lateral axis $x$ (along the relay surface) can be sampled at a lower rate, $\lambda_{sx}$, since the frequency content of the wavefront is bounded along this axis. Under the fresnel approximation, we find the ratio of these sampling intervals depends on the location of the hidden object $(x_0, z_0)$ relative to the relay wall $(x_r, z_r)$:

\begin{align}
\frac{\lambda_{sx}}{\lambda_{sz}} =  \frac{2|z_r - z_0|}{|x_r-x_0|} > D
\label{eq:upsample2}
\end{align}

We can use this expression to find the regimes $(|x_r - x_0|, |z_r-z_0|)$ for which this constraint satisfies some downsampling factor given by $D$ (Table \ref{table:Downsample}). Setting $D>1$ enables compression of the measurement.

In practice, it is easy to satisfy this criterion given by Eq \ref{eq:upsample2} since most hidden scenes are usually a few meters away from the relay surface to be out of the line of sight of the observer (Fig \ref{fig:Teaser}). We devise a simple algorithm to demonstrate the compressibility of the measurement. For a given measurement sampled at $\lambda_s$:

\begin{enumerate}[leftmargin=2\parindent]
    \item  Discard every $n^{\text{th}}$ sample. 
    \item  Interpolate back to the original grid with an algorithm of choice, such as nearest neighbors. 
    \item  Reconstruct using fast RSD algorithm, with the wavelength set to $2\lambda_s$
\end{enumerate}

We note that both the interpolation and reconstruction are performed using the temporal Fourier domain of the transient measurement, which is typically around ten times smaller than the time domain histograms due to the phasor field bandpass filter. Therefore, this method reconstructs on a dataset that has been compressed both spatially and temporally, and can be utilized to overcome data bandwidth issues that are a consequence of capturing billions of photons with an increasing amount of SPAD pixels.

 \newcommand{\addpicA}{\includegraphics[width=10em]{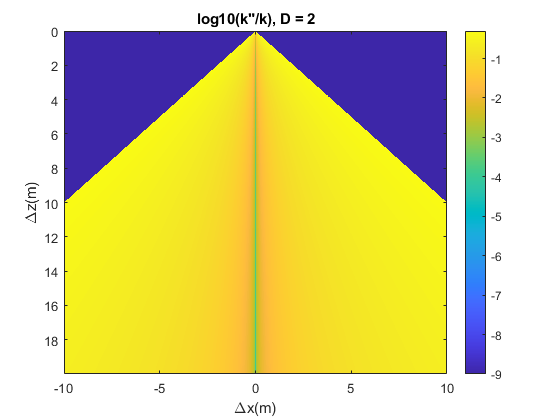}}
\newcommand{\addpicB}{\includegraphics[width=10em]{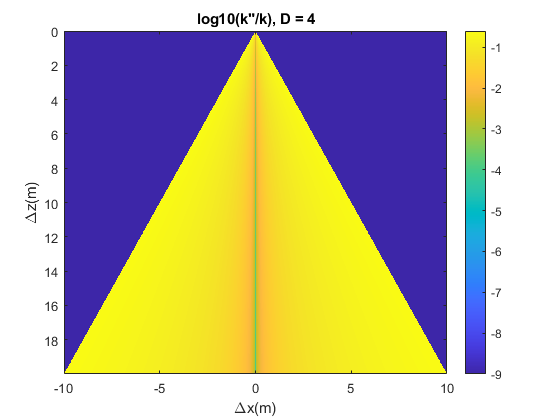}}

\newcolumntype{C}{>{\centering\arraybackslash}m{10em}}

\begin{table}[H]\sffamily
\centering
\begin{tabular}{l*2{C}@{}}
\toprule
\centering
 & D=2 & D=4 \\ 
\midrule
 & \addpicA & \addpicB \\
\bottomrule 
\end{tabular}
\caption{\centering
Yellow regime specifies the valid transverse offsets $x_r-x_0$ for a point source from the relay wall for a given downsampling rate D}
\label{table:Downsample}
\end{table} 

\section{Methods: Non-Uniform RSD}
\label{sec:NUFFT}

The Non-Uniform Fast Fourier Transform (NUFFT) is an efficient method for applying the Fast Fourier Transform (FFT) to signals sampled at non-uniform intervals. While many techniques exist in the literature to solve this issue \cite{bagchi_nonuniform_1999, shimobaba_nonuniform_2013}, our approach takes advantage of the spatial oversampling redundancy described in Section \ref{sec:NLOS_CS} to maintain the computational efficiency of the FFT. The key idea is to use interpolation schemes that approximate the non-uniformly sampled signal on a dense, uniform grid, allowing the FFT to compute the spectrum efficiently. Significant research has been devoted to designing these interpolation methods, enabling NUFFT implementations to achieve nearly $O(N\log N)$ complexity with low approximation error \cite{dutt_fast_1993, dutt_fast_1995, greengard_accelerating_2004, steidl_note_1998, barnett_parallel_2019, BARNETT20211, shih_cufinufft_2021, potts_uniform_2021, fessler_nonuniform_2003}.

We first demonstrate the basic idea of the NUFFT using a 2D signal, but this principle can be easily extended to 3D. Suppose we sample a 2D signal, $u(x, y)$, at irregular spacing. Then these sampled points can be compiled into a list denoted by $(x_\ell , y_\ell)$, where $\ell = 1, 2, \dots, \mathcal{L}$:

\begin{equation}
\begin{aligned}
\mathrm{u}[x_\ell , y_\ell] = \sum_{\ell=1}^{\mathcal{L}} \mathrm{u}(x, y)  \delta(x-x_\ell, y - y_\ell)
\end{aligned}
\end{equation}

In type I NUFFT (NUFFT-1), we synthesize a uniform dual spectrum from non-uniformly sampled points in the primal domain (Row 3, Fig. \ref{fig:MFFT_Summary}). Rescaling our list i.e. $(x_\ell, y_\ell) \in [-\pi, \pi)^2$ and assuming that the signal is periodic over $2\pi$, we can compute the Fourier series coefficients: 

\begin{equation}
\begin{split}
\mathrm{U}[m^{\prime} , n^{\prime}] = \sum_{\ell = 1}^{\mathcal{L}}  \mathrm{u}[x_\ell , y_\ell] e^{-j \left(x_\ell \cdot m^{\prime} + y_\ell \cdot n^{\prime}  \right)} \\
(m^{\prime} , n^{\prime})  \in \mathcal{I}_{M, N}
\end{split}
\end{equation}

where 

\begin{equation}
I_\mathcal{L} := \{ \ell^{\prime} \in \mathcal{Z}: -\mathcal{L}/2 \leq \ell < \mathcal{L}/2  \}
\end{equation}

refers to a 2D grid of Fourier frequencies. Type 2 NUFFT (NUFFT-2) is the adjoint of type 1 NUFFT. Using a uniform dual spectrum, it synthesizes a primal function at non-uniform samples (Row 4, Fig. \ref{fig:MFFT_Summary}). Given a regular grid of Fourier coefficients, $\mathrm{U}[m^{\prime} , n^{\prime}]$, we can evaluate the signal at non-uniform locations  $(x_\ell, y_\ell) \in [-\pi, \pi)^2$ with:

\begin{equation}
\begin{split}
\mathrm{u}[x_\ell , y_\ell] = \sum_{m^{\prime}=-M / 2}^{M / 2-1} \sum_{n^{\prime}=-N / 2}^{N / 2-1}  \mathrm{U}[m^{\prime} , n^{\prime}]   e^{j \left(x_\ell \cdot m^{\prime} + y_\ell \cdot n^{\prime}  \right)}, 
\\ \ell = 1, 2, \dots, \mathcal{L}
\end{split}
\end{equation}

In general, there are three steps for fast computation of NUFFT-1:

\begin{enumerate}
    \item Spreading or convolving each non-uniformly sampled point onto a dense grid using a blur kernel, $\psi(x)$.
    \item Using fast FFT to generate the Fourier coefficients of this blurred signal on the dense grid. 
    \item Deblurring or deconvolving by dividing the output pointwise by Fourier coefficients of the blur kernel.  
\end{enumerate}

Computing NUFFT-2 simply involves reversing these steps, where the final step involves interpolating the blurred signal to points with non-uniform spacing. 

\begin{figure*}[!htbp]
    \centering
\captionsetup{skip=3pt}\includegraphics[width=1\textwidth]{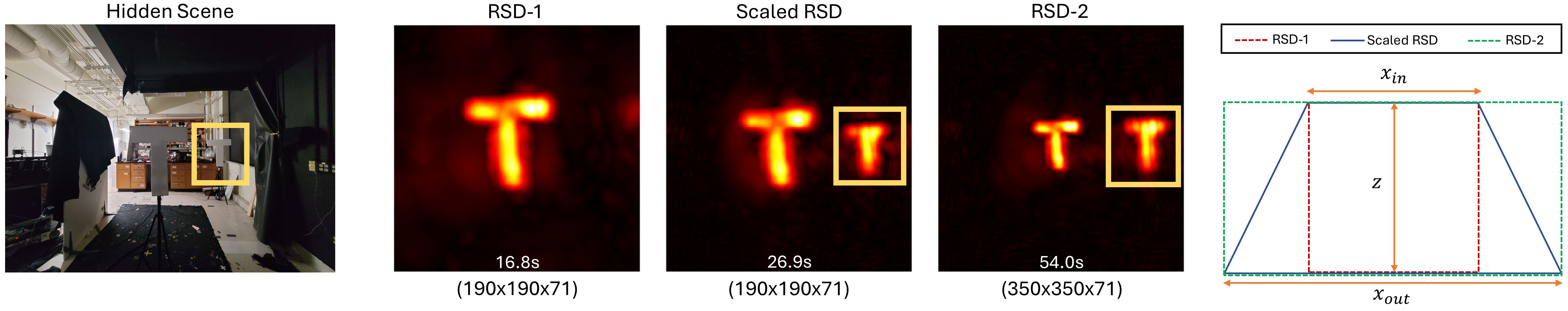}
    \caption{Column 1 presents the hidden scene as viewed from the relay surface. Columns 2, 3, and 4 display reconstructions using the standard RSD and the scaled RSD, where a max filter is applied along the depth to generate 2D images.The average reconstruction time in seconds is shown in white text at the bottom center of each image, while the black text below it indicates the dimension-wise number of voxels (X x Y x Z) used for the reconstruction volume. Column 5 provides a top view of the reconstructed volumes for each algorithm, illustrating that both the scaled RSD and RSD-2 reconstruct larger lateral areas. Moreover, the scaled RSD captures perspective effects that are absent in the standard RSD.}\label{fig:SRSD_Volume_OverSampled2}
\end{figure*}

\paragraph{Complexity} Let's assume we compute the NUFFT algorithm for a uniform d-dimensional grid of Fourier modes, given by $G: G_1 \times G_2 \times \dots G_d$. The computational complexity for the NUFFT algorithm has shown to be in the literature  \cite{dutt_fast_1993} to be $O(G \log G + \mathcal{L} \log^d (1/\epsilon))$, where $\mathcal{L}$ is the number of non-uniform samples, $d$ is the number of dimensions in the data, and $\epsilon$ is a user prescribed tolerance for accuracy. We set $G = N^d$ when $G$ is a d-dimensional cube with side length given by $N$. In this work, $\mathcal{L} \leq N^d$ so the computational complexity is given by $O(N^d \log N)$ where $d=2$ and $d=3$ for datasets collected on planar and non-planar relay surfaces respectively.

\paragraph{Accuracy} There has been extensive work on finding optimal blur kernels, $\psi(x)$, that allow for fast computation and interpolation while quantifying and minimizing approximation error that arises from using these kernels \cite{dutt_fast_1993, dutt_fast_1995, greengard_accelerating_2004, steidl_note_1998, barnett_parallel_2019, BARNETT20211, shih_cufinufft_2021, potts_uniform_2021, fessler_nonuniform_2003}. The exact error depends on the specific blurring kernel, and is negligble for this work since our imaging system already oversamples the relay surface. In this work, we utilize the prebuilt Matlab library titled \textit{nufftn} since we find its performance to be sufficient for our application.

\subsection{NURSD}
To utilize the NUFFT in the RSD, we define three new RSDs.

\paragraph{NURSD-1}

\begin{equation}
 \begin{split} 
\mathcal{P}_{\mathcal{F}} \left( m\Delta x, n\Delta y, z_v, \omega \right) 
&= \\
\operatorname{IFFT}\Bigl\{ \operatorname{NUFFT-1}\left\{\mathcal{P}_{\mathcal{F}} \left( x_\ell, y_\ell, 0, \omega \right) \right\}  & \cdot  \operatorname{FFT}\left\{G\left(m\Delta x, n\Delta y, z_v, \omega\right) \right\}\Bigr\}
\label{eq:NURSD1}
\end{split}   
\end{equation}

\paragraph{NURSD-2}

\begin{equation}
 \begin{split} 
\mathcal{P}_{\mathcal{F}} \left( x_\ell, y_\ell, z_v, \omega \right) 
&= \\
\operatorname{NUFFT-2}\Bigl\{ \operatorname{FFT}\left\{\mathcal{P}_{\mathcal{F}} \left( m\Delta x, n\Delta y, 0, \omega \right) \right\}  & \cdot  \operatorname{FFT}\left\{G\left(m\Delta x, n\Delta y, z_v, \omega\right) \right\}\Bigr\}
\label{eq:NURSD2}
\end{split}   
\end{equation}

\paragraph{NURSD-3}

\begin{equation}
 \begin{split} 
\mathcal{P}_{\mathcal{F}} \left( x_\ell, y_\ell, z_v, \omega \right) 
&= \\
\operatorname{NUFFT-2}\Bigl\{ \operatorname{NUFFT-1}\left\{\mathcal{P}_{\mathcal{F}} \left(x_\ell, y_\ell, 0, \omega \right) \right\}  & \cdot  \operatorname{FFT}\left\{G\left(m\Delta x, n\Delta y, z_v, \omega\right) \right\}\Bigr\}
\label{eq:NURSD3}
\end{split}   
\end{equation}

where the non uniform coordinates  $(x_\ell, y_\ell)$ are rescaled to lie in the range $[-\pi, \pi)^2$, and are arranged in an unordered list indexed by $\ell = 1, \dots \mathcal{L}$. NURSD-1 is able to reconstruct from non-uniform sampling scheme on the relay surface, and can be compared with the 3D RSD. NURSD-2 reconstructs at arbitrary locations at each reconstructed plane, further reducing the number of samples needed to store the reconstruction grid and may be useful to combine with optimization techniques which iteratively refine where to sample the voxel grid. Finally, NURSD-3 combines NURSD-1 and NURSD-2 and can be used for generalized sampling schemes for both acquisition and reconstruction.

\subsection{3D NURSD}
In stage 1, the 3D RSD approximates the input signal on a uniform grid, $\vec{x}_{\mathrm{c}}$, using Eq \ref{eq:3d_rsd_Int} followed by taking the spatial 3D FFT to propagate the virtual wavefront using convolution (see Eq \ref{eq:3d_rsd}). We simply replace these two steps with a 3D NUFFT-1, since the NUFFT has been developed and optimized for these two steps:

\paragraph{3D NURSD}

\begin{equation}
 \begin{split} 
\mathcal{P}_{\mathcal{F}} \left( m\Delta x, n\Delta y, z_v, \omega \right) 
&= \\
\operatorname{IFFT}\Bigl\{ \operatorname{NUFFT-1}\left\{\mathcal{P}_{\mathcal{F}} \left( x_\ell, y_\ell, z_\ell - z_0, \omega \right) \right\}  & \cdot  \operatorname{FFT}\left\{G\left(m\Delta x, n\Delta y, z_v, \omega\right) \right\}\Bigr\}
\label{eq:3D_NURSD}
\end{split}   
\end{equation}

\subsection{Novel Fusions}
In Supplement Section \ref{sec:Fusions}, we discuss novel fusions of the various algorithms introduced in this paper. For example, the NURSD-2 algorithm can be combined with the Scaled RSD to sample at arbitrary locations on the scaled grid.
\begin{figure}
    \centering
    \captionsetup{skip=3pt}
    \includegraphics[width=0.7\columnwidth]{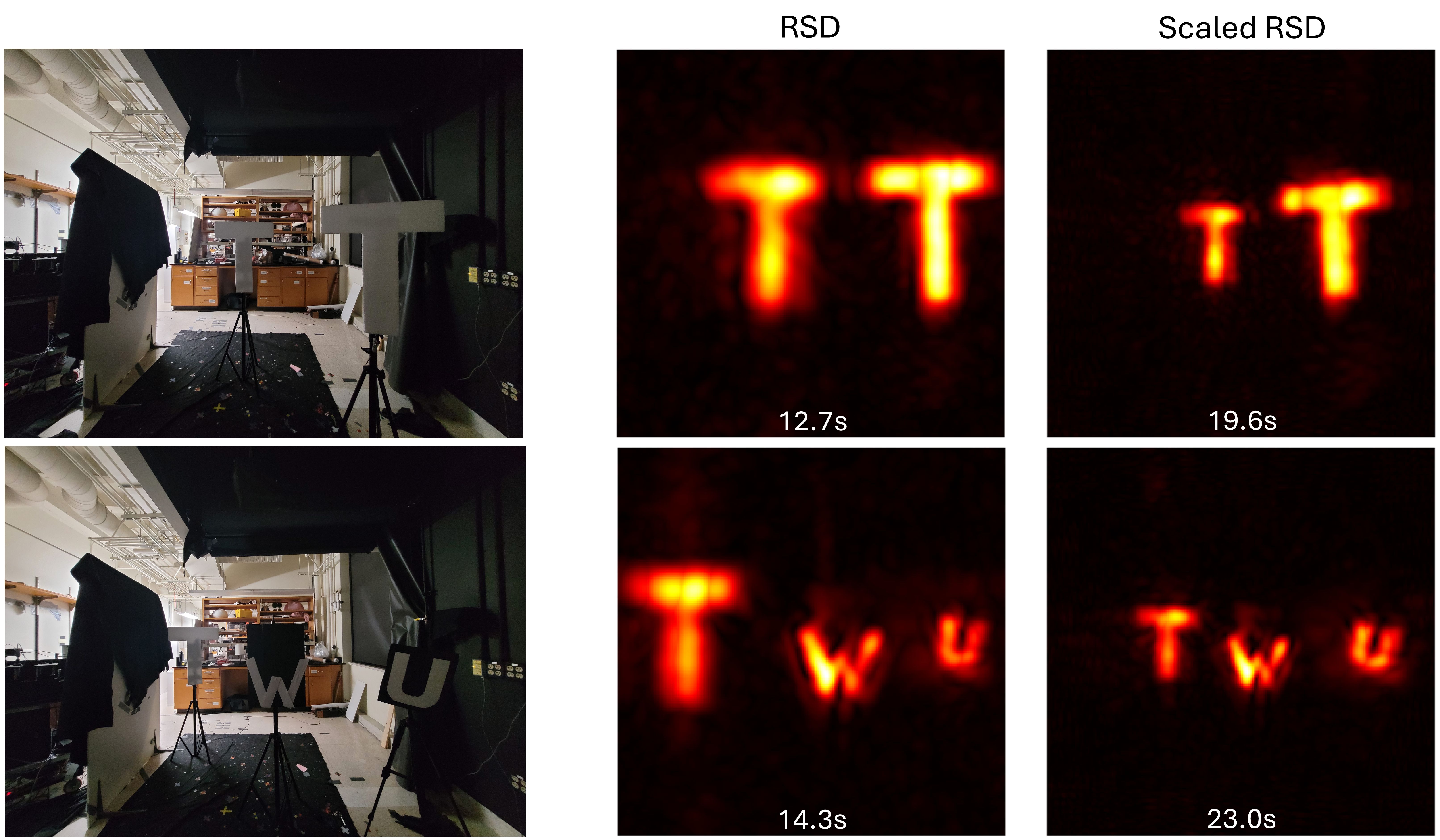}
    \caption{Column 1 shows the image of the hidden scene. Columns 2 and 3 display the standard RSD reconstruction and the scaled RSD reconstruction, respectively, after applying a max filter operation. Row 1: Two letter T's of the same size are shown, with the T in the middle appearing smaller in the SRSD reconstruction due to its greater depth. Row 2: Letters T, W, and U of different sizes are arranged at varying depths so that they appear the same size when viewed from the relay wall. The numbers at the bottom center indicate the time in seconds needed to reconstruct each image.}
    \label{fig:SRSD_Perspective}
\end{figure}

\section{Results}
\label{sec:Results}

\subsection{Scaled RSD}
We implemented the Scaled RSD (SRSD), and apply it to real data collected using our experimental setup, as illustrated in Fig. \ref{fig:Hardware}. In our reconstructions, we use the same scaling factor $\alpha$ for both lateral spatial dimensions. Additionally, our exposure time per spatial grid position is around 3ms, so we use five illumination positions $\vec{x_p}$ to enhance the signal-to-noise ratio (SNR) for the reconstruction. The average reconstruction time per illumination position, $\vec{x_p}$, is recorded at the bottom center of each reconstructed image. For the same number of samples/voxels in memory, the SRSD takes around 1.5x longer to reconstruct than the standard RSD due to three additional FFT computations. However, using the SRSD offers two key advantages. 

The first, as shown in Fig. \ref{fig:SRSD_Volume_OverSampled2}, is its ability to reconstruct larger volumes compared to the standard RSD. When the number of voxels is fixed to (190x190x71) voxels, the Scaled RSD reconstructs the letter "T" placed to the right (Column 3), while the standard does not (Column 2). To address this, the standard RSD can incorporate zero-padding to reconstruct a larger lateral area (Column 4), which captures the 'T' but requires (350×350×71) voxels and significantly more reconstruction time. In contrast, the scaled RSD achieves this with fewer voxels and nearly half the time.

The second benefit is that the reconstructed image matches human perspective, since objects located further away from the relay surface appear to be smaller. To visualize the 3D reconstruction using a 2D image, we display the voxel with maximum intensity along the depth axis. For the volume reconstructed with SRSD, this projection operation aligns with the \textit{perspective projection} used in computer graphics. In contrast, applying this operation to the volume reconstructed with standard RSD results in an \textit{orthographic projection}. Fig. \ref{fig:SRSD_Perspective} presents smartphone images of hidden scenes taken from the perspective of the relay wall (Column 1). Column 2 shows the standard RSD reconstruction with orthographic projection, while Column 3 demonstrates how SRSD reconstruction retains human (and smartphone) perspective, as the 'T' at a larger depth appears smaller in size (Row 1, Column 3). Row 2 depicts a scene with letters T, W, and U of different sizes placed at varying depths so that they appear to be of the same size (Row 2, Column 1). The SRSD reconstruction (Row 2, Column 3) accurately captures this perspective, while the RSD reconstruction displays the letters in their actual sizes (Row 2, Column 2).

\subsection{SubSampling}
\label{sec:SubSampling}
\begin{figure}
    \centering
    \captionsetup{skip=3pt}
    \includegraphics[width=1\columnwidth]{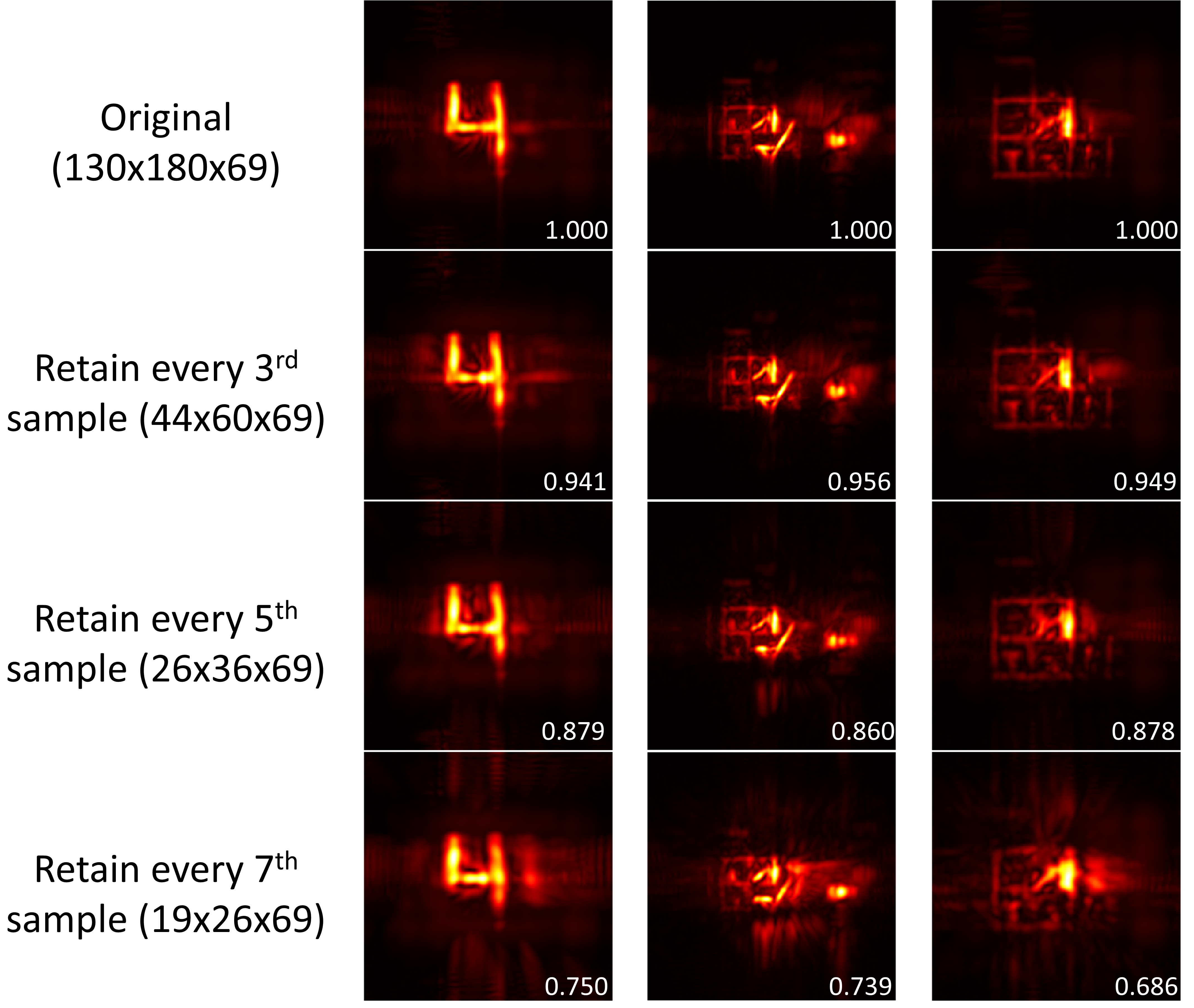}
    \caption{Standard RSD reconstruction results after spatial downsampling and nearest neighbor interpolation to original dimensions. We note that the office scene uniquely has more frequency components with dimensions (130x180x139) since this scene has a larger depth range. SSIM score is shown on the bottom right, and is computed with the reference set to the reconstruction in the top row.}
    \label{fig:upsample_NN}
\end{figure}

We use publicly available datasets \cite{liu_non-line--sight_2019, liu_non-line--sight_2021} to showcase our results. The datasets were collected using 1 cm spacing on a 1.8 m x 1.3 m relay surface, and reconstructed with $\lambda_c = 4$ cm. Discarding 4 out of 5 samples is equivalent to collecting the dataset with 5 cm spacing. Fig. \ref{fig:upsample_NN} demonstrates that interpolating this compressed, 5 cm dataset using nearest neighbor scheme does not significantly degrade reconstruction quality. To quantify this, we compute a structural similarity index measure (SSIM) score where the reference is set to the reconstruction generated with the original dataset where no samples are removed (row 1 of Fig. \ref{fig:upsample_NN}). Using a linear interpolation scheme shows marginal improvement, as shown in Fig. \ref{fig:upsample_Linear}. 

\subsubsection{Impact of Noise}
Our derivation for Eq \ref{eq:upsample} does not consider the impact of photon noise, which becomes important under low exposure times. We show in Fig. \ref{fig:upsample_noise} how reducing the acquisition times significantly degrades the reconstruction quality using our simple interpolation scheme. Specifically, the speckle artifacts are greatly exacerbated when interpolating noisy measurements. Our naive interpolation can be replaced or augmented with smart, denoising algorithms \cite{cho_learning_2024, pan_onsite_2022} that account for this quantization noise prior to reconstruction.

\subsubsection{Phasor Field Filtering} 
Alternately, we can increase $\lambda_c$ in the phasor field kernel to denoise the reconstruction or increase data compression by discarding additional samples, but this comes at the expense of reconstruction quality (Fig.  \ref{fig:upsample_wavelength}). Conversely, we can reconstruct the hidden scene at higher resolutions by reducing the phasor field wavelength, and simultaneously upsampling the original dataset. This generates slightly cleaner reconstructions, and reveals additional details in the scene (Fig. \ref{fig:upsample_HR}). Since we don't have ground truth reference, it does not make sense to compute an SSIM score to quantify the improvement here.

\subsubsection{Data Savings}
For a $N^2$ grid of spatial positions and $T$ timebins, the number of data points in the transient measurement is usually $N^2T$. Since the interpolation and the RSD reconstruction operate directly on the frequency domain of the data, we can store the filtered complex phasor field fourier coefficients, \cite{ liu_phasor_2020, nam_low-latency_2021, gutierrez-barragan_compressive_2022} giving a dataset of size $N^2(2F)$, where $F$ is the number of frequency coefficients and factor of 2 is needed to store the real and imaginary components of the complex coefficients. Let $N^\prime = N/D$ to represent the number of spatial points in the subsampled grid. For $D = 5$, and since $F \approx T/10$, we get a total data compression of $N^2T/((N^\prime)^2 2F) \approx 125)$. As SPAD arrays gain more pixels, this compression is useful for reducing the large memory footprint when capturing the full SPAD and laser grids simultaneously.

\twocolumn

\begin{figure}
    \centering
    \captionsetup{skip=3pt}
    \includegraphics[width=1\columnwidth]{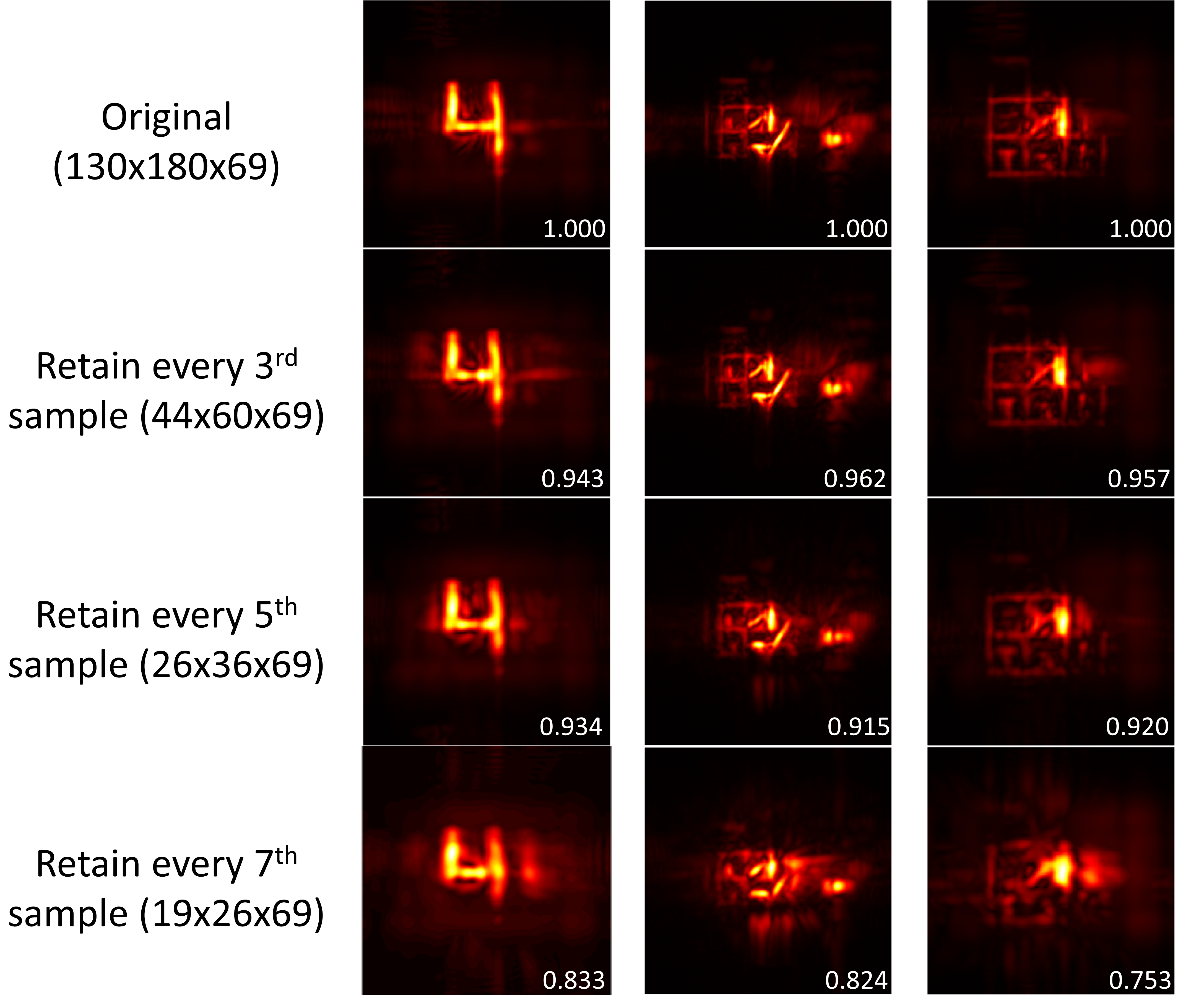}
    \caption{Standard RSD reconstruction results after spatial downsampling and linear interpolation to original dimensions. We note that the office scene uniquely has more frequency components with original dimensions (130x180x139) since this scene has a larger depth range. SSIM score is shown on the bottom right, and is computed with the reference set to the reconstruction in the top row.}
    \label{fig:upsample_Linear}
\end{figure}

\begin{figure}
    \centering
    \captionsetup{skip=3pt}    \includegraphics[width=1\columnwidth]{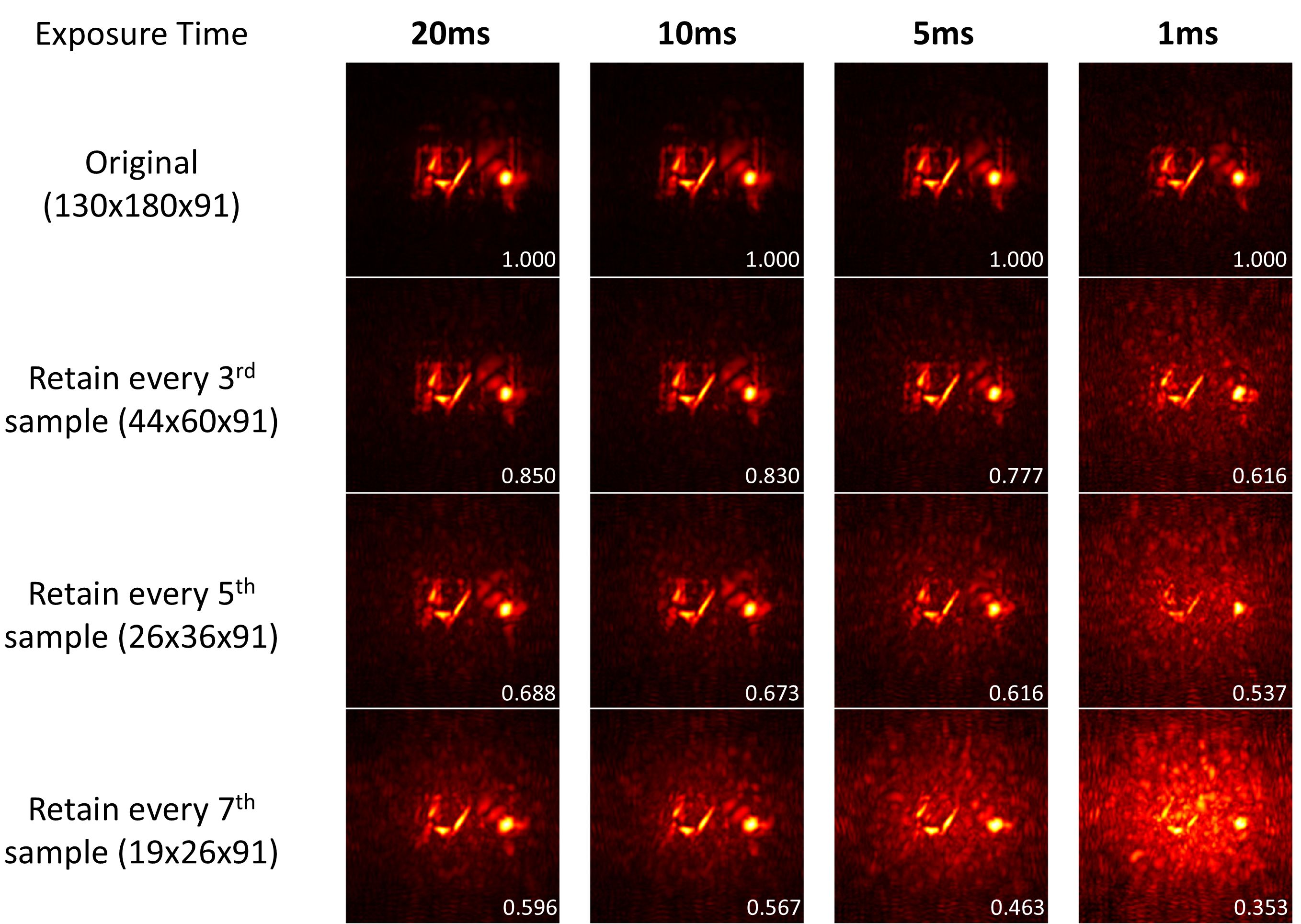}
    \caption{RSD reconstruction results after spatial downsampling and nearest neighbor interpolation to original dimensions for different exposure times for the office scene. The listed exposure time is for each spatial grid position. We note that the office scene uniquely has more frequency components with original dimensions (130x180x139). SSIM score is shown on the bottom right, and is computed with the reconstruction in the top row set to the reference.}
    \label{fig:upsample_noise}
\end{figure}

\begin{figure}
    \centering
    \captionsetup{skip=3pt}
\includegraphics[width=1\columnwidth]{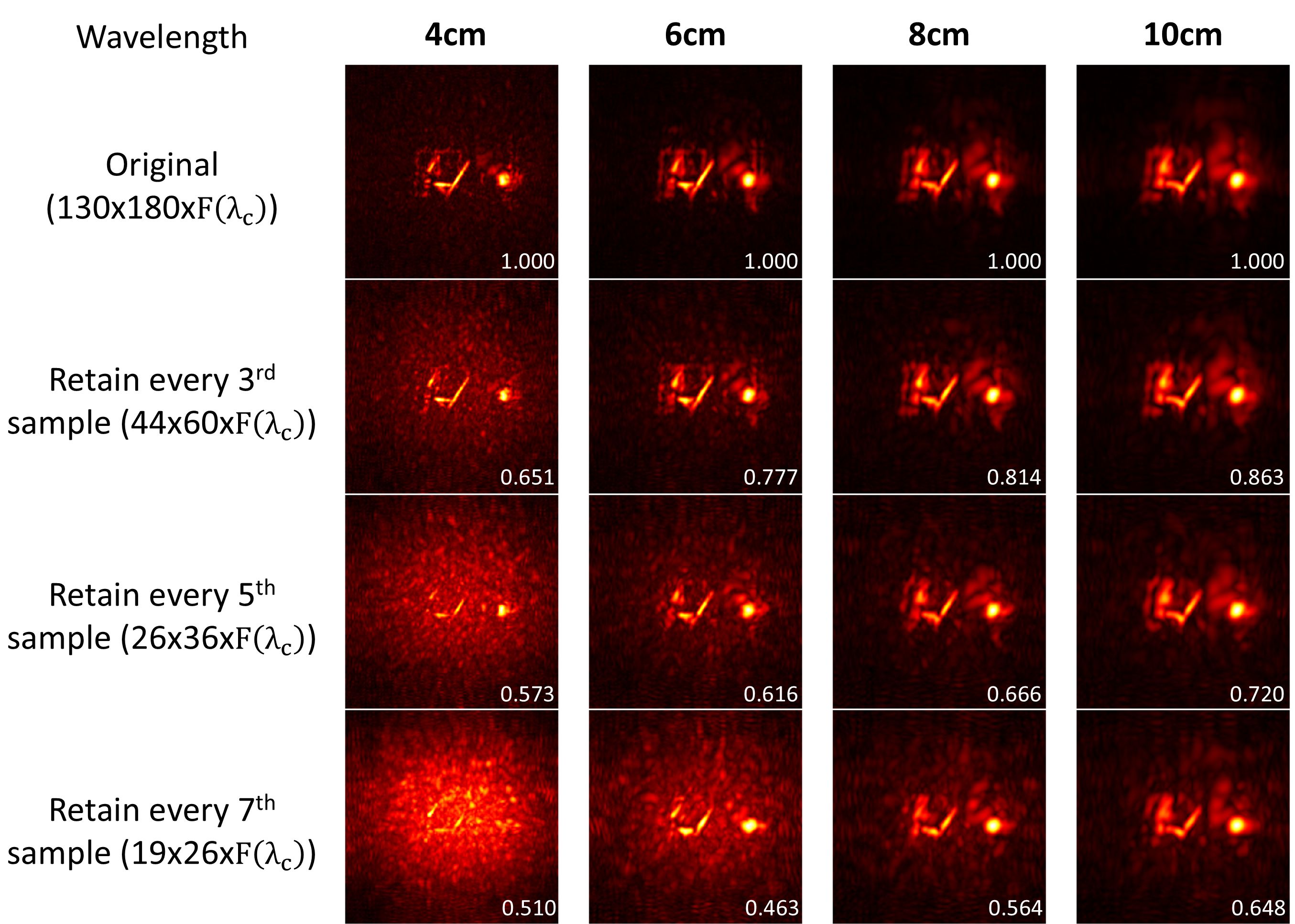}
    \caption{RSD reconstruction results after spatial downsampling and nearest neighbor interpolation to original dimensions for different wavelengths, $\lambda_c$, for the 5ms office scene. The number of frequency components used in the reconstructions, $F$, depends on the wavelength. SSIM score is shown on the bottom right, and is computed with the reconstruction in the top row set to the reference.}
    \label{fig:upsample_wavelength}
\end{figure}

\onecolumn

\begin{figure}
    \centering
    \captionsetup{skip=3pt}
    \includegraphics[width=0.5\columnwidth]{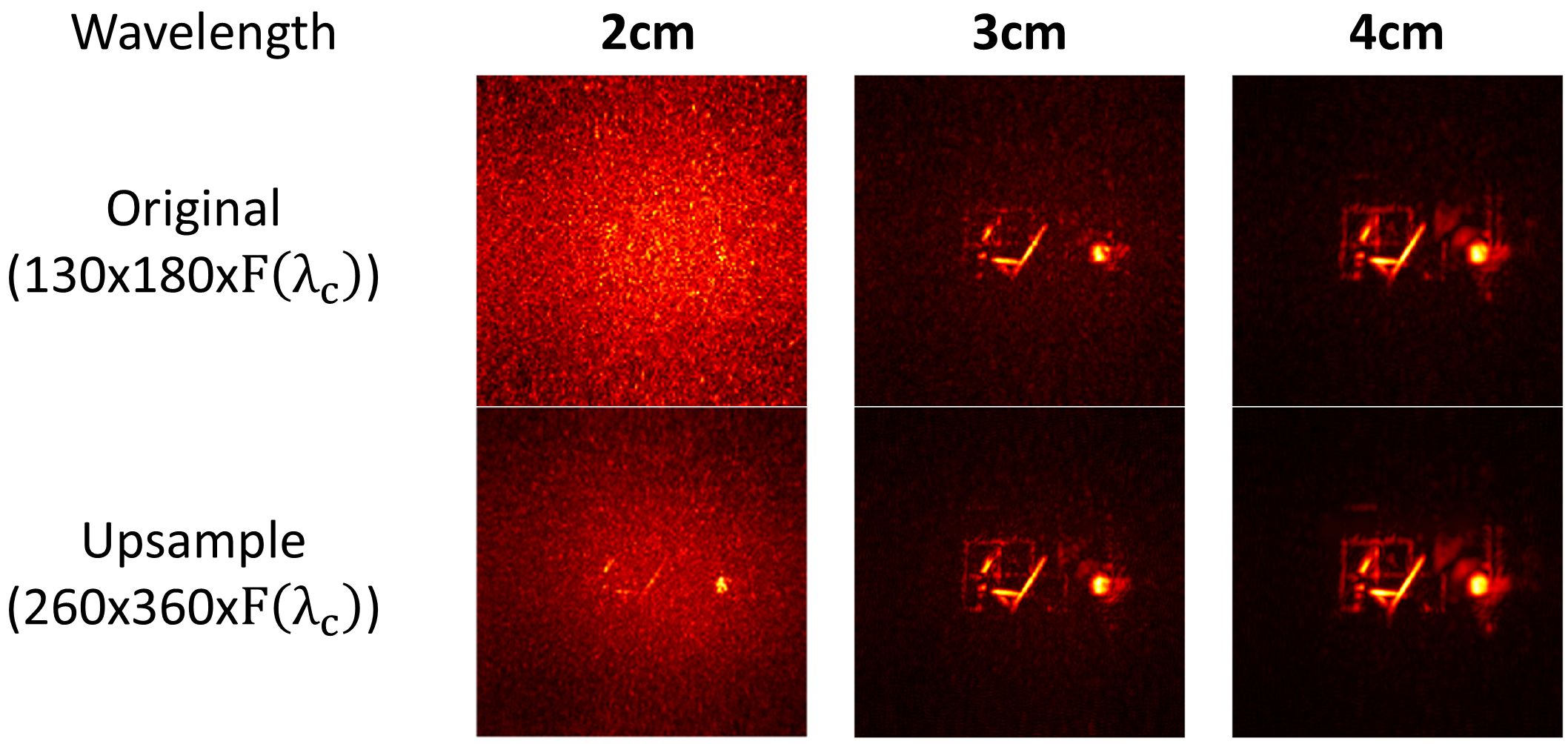}
    \caption{RSD reconstruction results after spatially upsampling the original dataset using  linear interpolation for different wavelengths, $\lambda_c$, for the 20ms office scene. This reveals previously missing details and slightly denoises the signal. The number of frequency components used in the reconstructions, $F$, depends on the wavelength.}
    \label{fig:upsample_HR}
\end{figure}
\section{Results: Non-Uniform RSD}
\label{sec:NURSD_Results}
We implement the NURSD and demonstrate four distinct applications using the three variants discussed in Section \ref{sec:NUFFT}. For generating the 2D and 3D NUFFT, we use the \textit{nufftn} function in MATLAB. Although faster implementations with optimized blur kernels are available \cite{barnett_parallel_2019}, we find the performance of the MATLAB implementation satisfactory for our purposes. To ensure optimal performance in MATLAB, which requires input sample locations to be positive integers, we shift the relay wall positions to positive values and round them to the nearest centimeter. The nufftn-based Matlab reconstruction is slightly shifted relative to the RSD reconstruction, with the exact shift varying slightly between reconstructions. To compute the SSIM score and quantify similarity between the \textit{nufftn} reconstructions and RSD (and 3D RSD) reconstructions, we first align the images by calculating the 2D correlation coefficient and then use the maximum value to determine the exact shift.

\subsection{Non-Uniform 2D Acquisition}
\begin{figure}[!t]
    \centering
    \captionsetup{skip=3pt}
    \includegraphics[width=0.7\columnwidth]{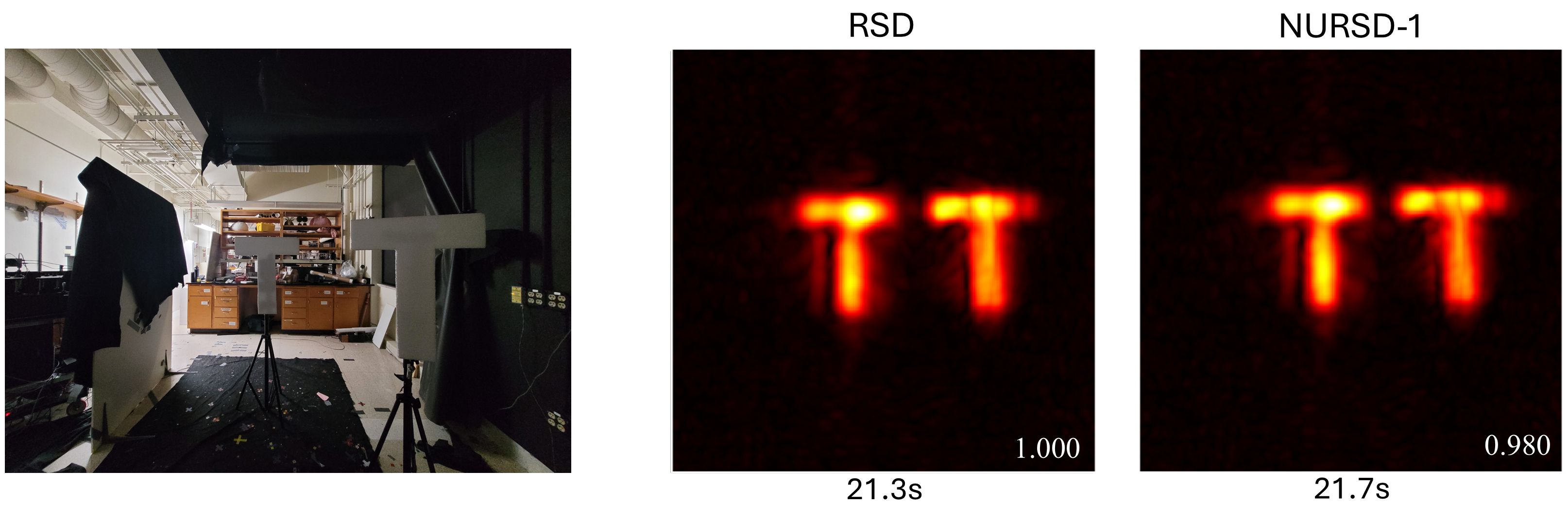}
    \caption{Comparison of reconstruction between the RSD and the NURSD-1 algorithms when the full grid is sampled on the relay surface. SSIM score is shown on the bottom right, and is computed with the RSD reconstruction as the reference.}
    \label{fig:NURSD1_vs_RSD}
\end{figure}

When the dataset is sampled non-uniformly on a planar relay wall, we can apply a 1 Stage, plane to plane NURSD-1 reconstruction algorithm described in Eq \ref{eq:NURSD1}. We first demonstrate the equivalence of the two algorithms by showing that, when the entire relay wall is sampled uniformly, they produce nearly identical results, as evidenced by the SSIM score close to 1 in Fig. \ref{fig:NURSD1_vs_RSD}. The runtime of both algorithms is very similar because the \textit{nufftn} computation is performed only once for each phasor field frequency, with the FFT-based propagation to every depth plane dominating the overall computation time.

\begin{figure}
    \centering
    \captionsetup{skip=3pt}
\includegraphics[width=0.7\columnwidth]{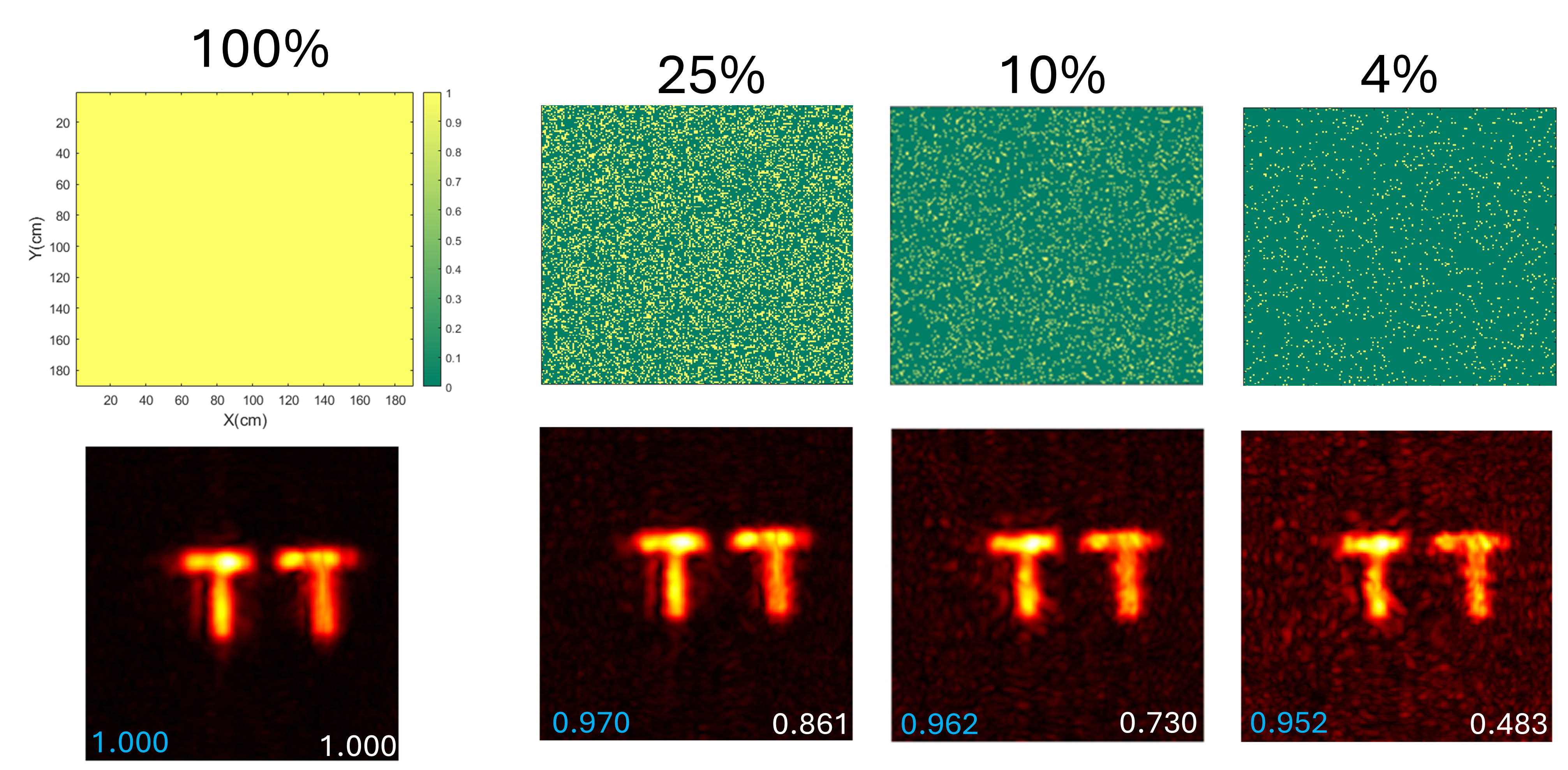}
    \caption{Top row shows the relay wall mask applied to the dataset. The NURSD enables arbitrary spatial sampling of the relay surface. SSIM score is shown on the bottom right, and is computed with the reconstruction in the top row set to the reference. Blue SSIM score shown after applying a low pass filter to remove the background}
    \label{fig:NURSD1_SubSample}
\end{figure}

Next, we use the NURSD-1 algorithm to reconstruct datasets collected with non-uniform spatial grids. We randomly subsample different percentages of a uniform spatial grid on the relay surface,  then reconstruct the data using the NURSD-1 algorithm, as shown in Fig. \ref{fig:NURSD1_SubSample}. Column 1 displays the NURSD-1 reconstruction using the entire spatial grid, which we use as the reference reconstruction for computing the SSIM score. The next three columns show reconstructions using progressively smaller percentages of the grid. The white numbers in each image show the SSIM score,which decreases as more positions are discarded. However, this degradation is primarily due to increased background noise since SNR reduces with fewer sampled positions. If we filter out the background using a max filter (by thresholding out the bottom 30 percent), then the SSIM score, shown in blue, remains close to 1 even when 96$\%$ of positions have been discarded. This reinforces our finding from Section \ref{sec:NLOS_CS} that existing NLOS setups oversample the relay surface.

\subsubsection{Detector Arrays}  
We collect a dataset with a 16 x 16 SPAD array (with 216 active pixels) focused over a large area on a planar relay surface. The focusing optics and the presence of hot pixels in the array means that the spatial detection grid on the relay surface is non-uniform (Column 2, Fig. \ref{fig:Hardware}). Fig. \ref{fig:NURSD1_SPADGRID_SUPP} in Supplement section \ref{sec:NURSD_SPADArray} demonstrates that the reconstructions generated using the NURSD-1 and FBP algorithms are consistent with each other.

\begin{figure}
    \centering
    \captionsetup{skip=3pt}
    \includegraphics[width=0.7\columnwidth]{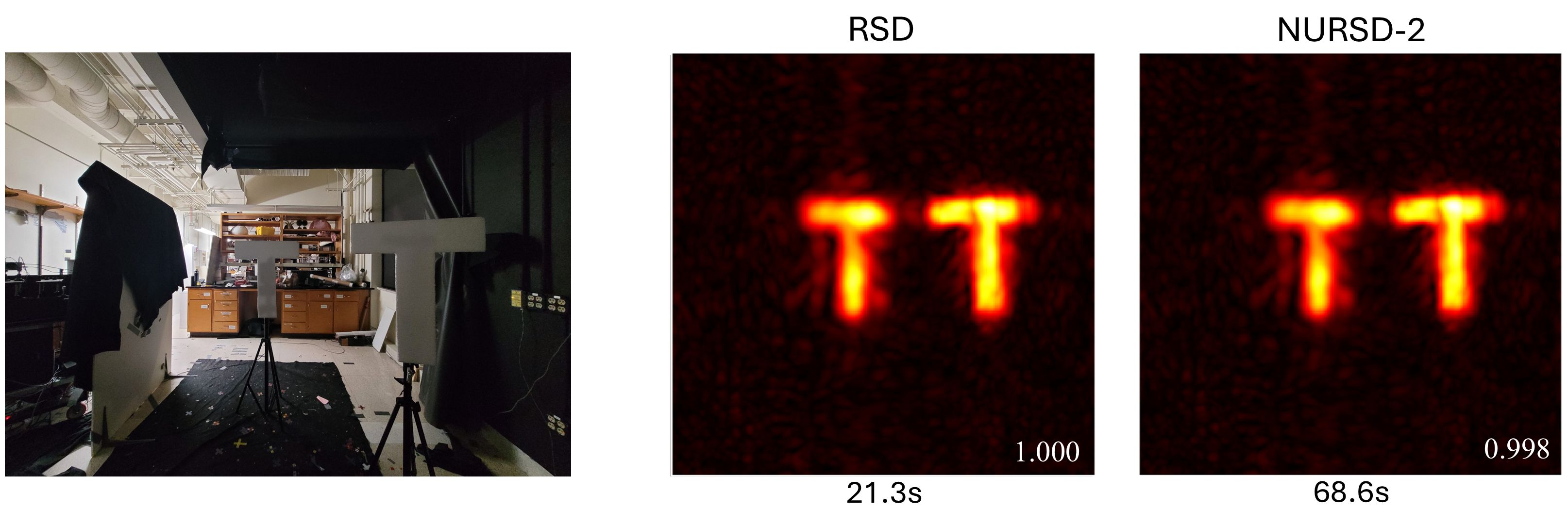}
    \caption{Comparison of reconstruction between the RSD and the NURSD-2 algorithms when the full voxel grid is sampled. SSIM score is shown on the bottom right, and is computed with the RSD reconstruction as the reference.}
    \label{fig:NURSD2_vs_RSD}
\end{figure}

\subsection{Non-Uniform Reconstruction}
\begin{figure}
    \centering
    \captionsetup{skip=3pt}
    \includegraphics[width=0.7\columnwidth]{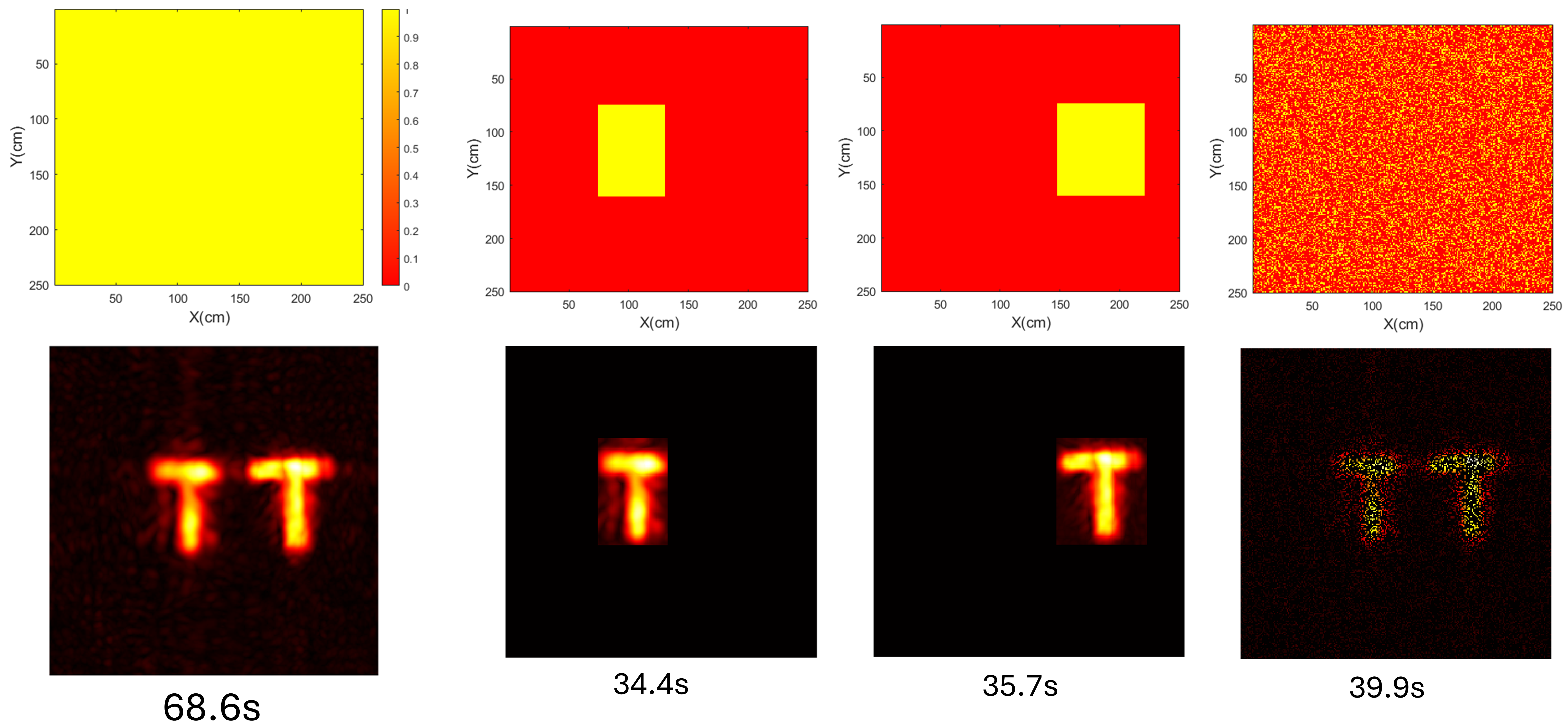}
    \caption{Top row shows the reconstruction mask for each plane in the reconstruction volume. The NURSD enables arbitrary spatial sampling of the voxel grid. Bottom row shows the corresponding reconstruction, along with computation time.}
    \label{fig:NURSD2_SubSample}
\end{figure}

We can utilize the NURSD-2 (Eq \ref{eq:NURSD2}) algorithm to sample the reconstruction volume arbitrarily while preserving the computational complexity. Fig. \ref{fig:NURSD2_vs_RSD} shows that the reconstruction quality is very similar to the RSD output, when the same sampling scheme is used for the reconstruction volume. In Fig.  \ref{fig:NURSD2_SubSample}, we reconstruct the hidden scene for different reconstruction masks. While the mask can be customized for each plane, we use the same mask for each plane here. We show the masks isolating each of the individual "T" letters, as well as a different mask where we sample 25$\%$ of available pixels in the mask (column 4). 

The NURSD-2 algorithm is slower than the NURSD-1 algorithm and standard RSD - this is because the \textit{nufftn} is called for each plane and for each frequency, while NURSD-1 calls the \textit{nufftn} function only once for each frequency. The NURSD-2 reconstruction time and memory requirements for the voxel grid decrease proportionally with the number of samples in the reconstruction mask (Fig. \ref{fig:NURSD1_vs_RSD}). Therefore, NURSD-2 can replace the standard RSD if operating under memory constraints or when a subset of the voxel grid must be sampled with high resolution - for example, when using optimization techniques to iteratively sample finer voxel grids. The matlab-based nufftn may be replaced with faster implementations  \cite{barnett_efficient_2021} to further reduce runtime.

\subsection{Non-Uniform Acquisition and Reconstruction}
\begin{figure}
    \centering
    \captionsetup{skip=3pt}
    \includegraphics[width=0.7\columnwidth]{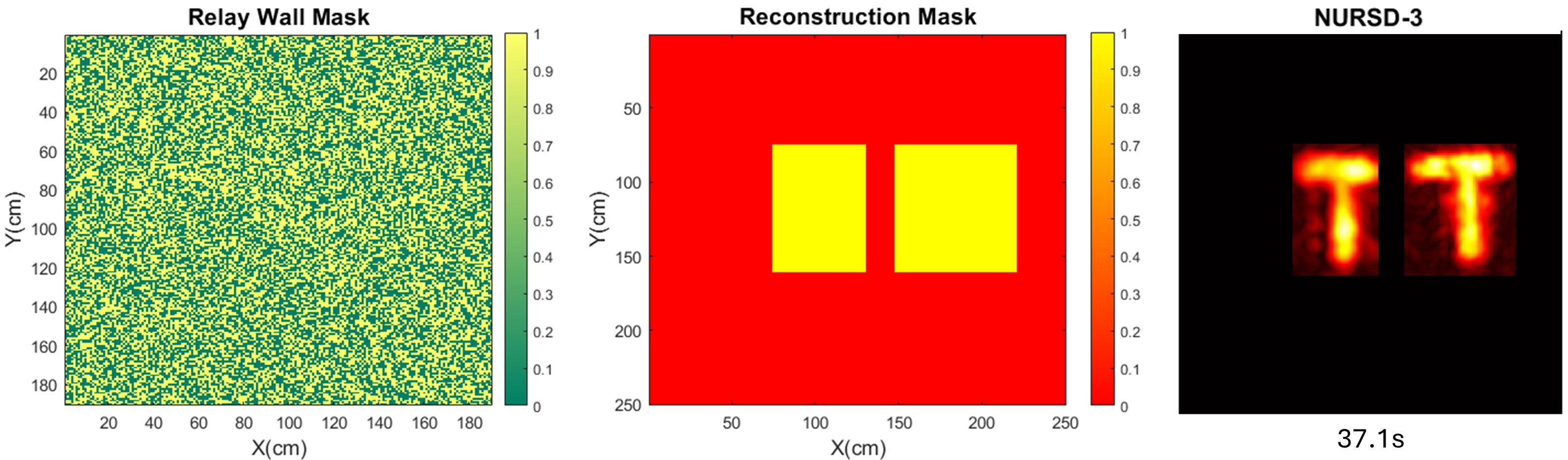}
    \caption{Columns 1 and 2 show the spatial masks for the relay surface and the reconstruction plane respectively. Column 3 shows the reconstruction, along with the computation time. This demonstrates that the NURSD enables arbitrary sampling of both the relay surface and the voxel grid.}
    \label{fig:NURSD3}
\end{figure}

Finally, the NURSD-3 algorithm in Eq \ref{eq:NURSD3} combines the previous two algorithms. We showcase this in Fig.  \ref{fig:NURSD3} by randomly subsampling 50$\%$ of positions on the relay surface (Column 1). Then we create a mask around the two Ts for reconstruction, shown in Column 2, while Column 3 shows the reconstruction. The reconstruction time is similar to that of the NURSD-2, since \textit{nufftn} function is called for every plane and every frequency as before. However, this is still orders of magnitude faster than the FBP algorithm which allows similar flexibility. 

\subsection{Non-Planar Acquisition}
 In this section, we apply the 3D NURSD algorithm to reconstruct NLOS datasets collected using non-planar relay surfaces, and compare the performance with the 3D RSD and the FBP algorithms. We replace Stage 1 of the 3D RSD algorithm, which consists of an interpolation step followed by 3D FFT, with a single 3D NUFFT. The FBP reconstructs the data sequentially for every grid position, and is very slow. However, there is no interpolation or approximation so the FBP serves as the gold standard in terms of reconstruction quality. Therefore, we compute an SSIM score where the reference is the FBP reconstruction.
 
\begin{figure*}
    \centering
    \captionsetup{skip=3pt}
    \includegraphics[width=1\columnwidth]{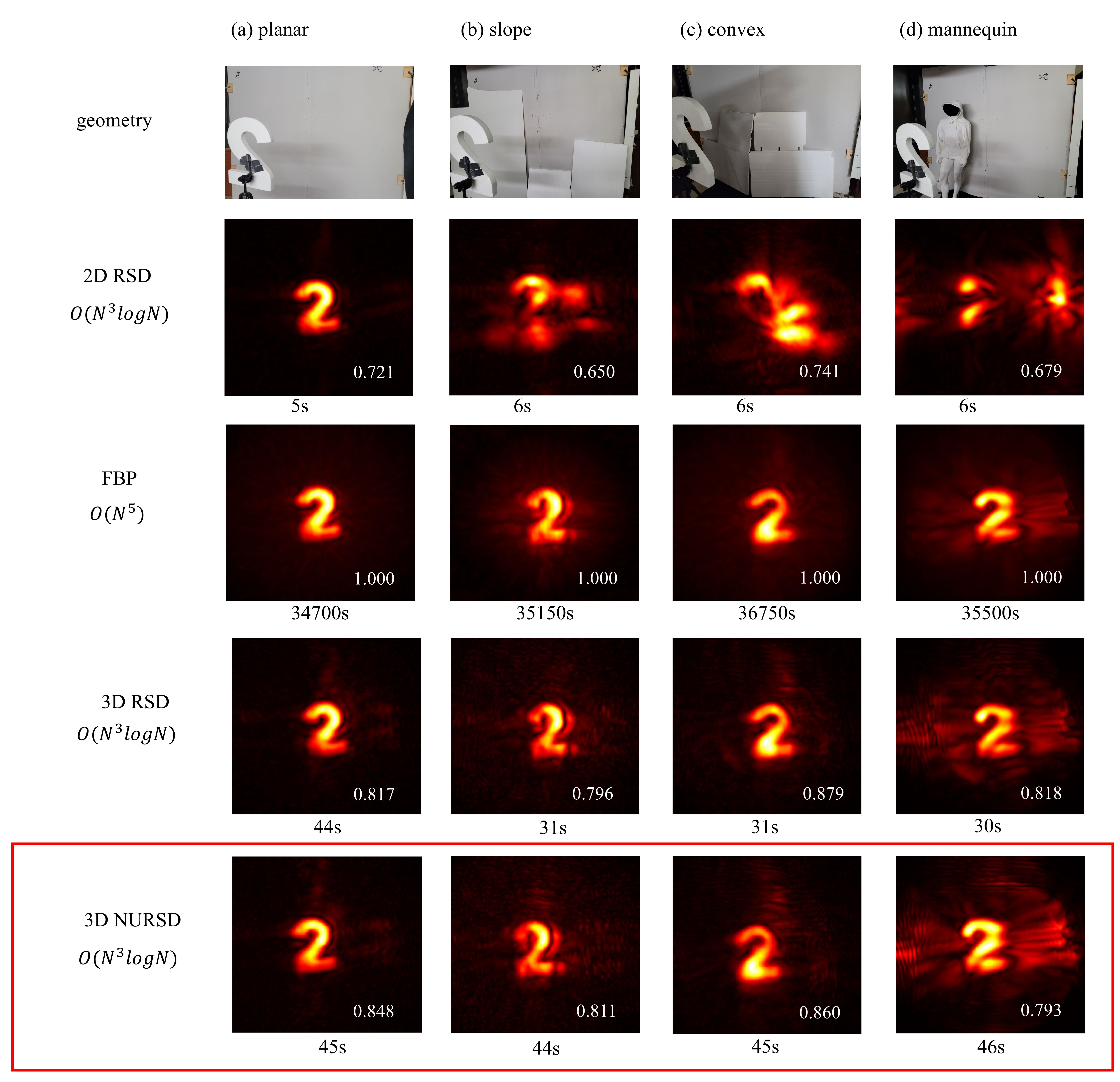}
    \caption{We compare the performance of RSD, FBP, 3D RSD, and 3D NURSD in rows 2, 3, 4, and 5 respectively for datasets acquired on planar (Column 1) and non-planar (Column 2 ,3, 4) relay surfaces for the same hidden scene. The performance of 3D RSD and 3D NURSD is comparable, and is order of magnitudes faster than FBP while maintaining comparable reconstruction quality. SSIM score is shown on the bottom right, and is computed with the FBP reconstruction set to the reference.}
    \label{fig:3D_NUFFT}
\end{figure*}

 As shown in Fig.  \ref{fig:3D_NUFFT}, the performance of the 3D RSD and 3D NURSD is fairly similar. Both are orders of magnitude faster than FBP and preserve the reconstruction quality, as shown by the relatively high SSIM scores. Qualitatively, the 3D NURSD reconstructions seem to reconstruct cleaner twos relative to the 3D RSD, but also seem to enhance background artifacts. Additionally, the 3D RSD is faster on average, this is because the algorithm interpolates only once to a uniform 3D grid, while the \textit{nufftn} function repeats the interpolation step for every frequency. On the other hand, the 3D RSD is memory intensive as it stores a uniform 3D grid for each frequency i.e. a 4D grid. As an additional benefit, mathematical guarantees for specific implementations for the 3D NUFFT, and thus the 3D NURSD can be found in the literature.

Since the Stage 2 is simply the plane to plane RSD algorithm, we can substitute in either the Scaled RSD algorithm (Eq \ref{eq:SRSD_conv}) or the NURSD-2 algorithm (Eq \ref{eq:NURSD2}) to reconstruct on a scaled grid or a grid with non-uniform spacing respectively.

\section{Discussion}
\label{sec:discussion}

\paragraph{Detector Arrays} Existing arrays have a limited number of pixels, which constrains how much these arrays can reduce acquisition times. In this work, we show that focusing the array over a large area on the relay surface, collecting photons for 20 - 50 ms exposure, and then interpolating will generate sufficiently good reconstructions. Row 3 of Fig. \ref{fig:upsample_Linear} shows that even a 32 x 32 pixel SPAD array can be used to generate high quality reconstructions. 

\paragraph{Poisson Noise} In this work, we demonstrate that simple interpolation on undersampled measurements fails to generate reconstructions of acceptable quality when each individual grid position is collected using a small exposure time. In this scenario, Poisson noise dominates the measurement which the interpolation can not account for. 
Existing deep learning approaches that learn to denoise this photon noise \cite{cho_learning_2024} in the measurement can serve as a useful replacement or supplement for this interpolation step, and improve reconstruction quality without increasing acquisition time.

\paragraph{Oversampling} Reconstruction algorithms developed for ToF NLOS tend to filter out high frequency information in the measurement and extract the low frequency information that preserves useful information about the hidden scene. We have shown that ToF NLOS imaging systems tend to oversample the relay surface, which allows us to compress the NLOS measurement spatially. This provides intuition for why previous work is able to generate reasonable reconstructions with spatial resampling \cite{nam_low-latency_2021}, and interpolation \cite{gu_fast_2023}. The normal move-out correction that converts non-confocal acquisition to confocal acquisition scheme is another spatial modification of the measurement that works better when the measurement is acquired with long exposure times \cite{liu_phasor_2020}. Our other finding is that the baseline sampling rate for confocal acquisition is double that of the confocal acquisition (see Supplement Section \ref{sec:NLOS_CS_Proof}). This partly explains why discarding the same number of spatial samples degrades the reconstruction quality of confocal datasets more than non-confocal ones \cite{liu_non-line--sight_2023, wang_non-line--sight_2023, cho_learning_2024}.

\paragraph{Relay Wall Calibration} This oversampling means that the NUFFT can be used directly on non-uniformly spaced spatial grid on the relay surface without compromising reconstruction quality. Therefore, the relay surface need not be calibrated to scan a uniform grid for either the illumination or the detection positions. 

\section{Conclusion}
\label{sec:conclusion}
In this work, we introduced the Scaled Rayleigh-Sommerfeld Diffraction (SRSD) and Non-Uniform Rayleigh-Sommerfeld Diffraction (NURSD) algorithms to address key challenges in NLOS imaging. These algorithms enable flexible and optimized sample schemes for the input and output while maintaining the computational complexity of existing state of the art algorithms. The NURSD excels in reconstructing datasets collected with non-uniform sampling on planar and non-planar relay surfaces with minimal loss in quality, alleviating the need to calibrate the relay surface to generate uniform grids. The SRSD generates reconstruction with optimized sampling for the voxel grid, and is particularly effective for reconstructing large-scale hidden volumes. These contributions provide a foundation for improving NLOS imaging in practical applications where efficiency and adaptability are critical. Future work will explore interpolation techniques that incorporate photon noise in the measurement or develop adaptive sampling frameworks that that dynamically adjust where to sample based on object motion or location. 

\section*{Acknowledgments}
This work was supported by the Air Force Office for Scientific Research (FA9550-21-1-0341). PP’s contribution is supported by the US Office of Naval Research under award No. N00014-21-1-2469 and by the US Joint Directed Energy Transition Office (JDETO). The authors acknowledge Simone Riccardo, Alberto Tosi, and their group at Politecnico di Milano for developing the 16 x 16 gated SPAD array, specifically designed for NLOS imaging applications, which was used to collect experimental data in this study.

\small
\bibliographystyle{IEEEtran}
\bibliography{references_T, bibliography}

\clearpage
\setcounter{page}{1}
\setcounter{section}{0}

{
\newpage
    \onecolumn
    \centering
    \Large
    \textbf{Optimized Sampling for Non-Line-of-Sight Imaging Using Modified Fast Fourier Transforms} \\
    \vspace{0.5em}Supplementary Material \\
    \vspace{1.0em}
}

\appendix
\appendix
In Section \ref{sec:Appendices_Math}, we provide the mathematical derivations referenced in the main paper. Section \ref{sec:Hardware} details the hardware setup for NLOS imaging used to collect the experimental datasets. In Section \ref{sec:Supp_Results}, we introduce novel fusions of existing algorithms and present additional results.

\section{Derivations}
\label{sec:Appendices_Math}

\subsection{Scaled FFT Derivation}
\label{sec:SFFT_Math}

Starting with the DFT equation: 

\begin{equation}
\begin{aligned}
\mathrm{U}\left[m^{\prime}, n^{\prime}\right] 
&\triangleq \sum_{m=-M / 2}^{M / 2-1} \sum_{n=-N / 2}^{N / 2-1} \mathrm{u}[m, n] \cdot  k_M^{- m \cdot m^{\prime}}  k_N^{- n \cdot n^{\prime}} 
\end{aligned}
\label{eq:2D_DFT_0}
\end{equation}

where 

\begin{equation}
  \begin{split}
 k_\mathcal{L}=\exp \left[-j \frac{2 \pi}{\mathcal{L}} \right]
\end{split}
\end{equation}

For the rest of the derivation, we use $(m, n)$ to represent the \textbf{integer} coordinates in the input space with $\delta_{\text{in}} = \delta$ sampling interval, and $(m^{\prime}, n^{\prime})$ to represent the \textbf{integer} coordinates in the output space with $\delta_{\text{in}} = \delta$ sampling rate. Note that $(m, n, m^{\prime}, n^{\prime})$ are all variables that range from $-N/2$ to $N/2-1$.

\sloppy For a given frequency, we simplify the input phasor field $\mathcal{P}_{\mathcal{F}} \left( x_c, y_c, 0, \omega \right)$ to a discrete 2D representation $\mathrm{u}[m, n]$, and the goal is to compute its Fourier transform on a scaled grid, i.e., $\mathrm{U}[m^{\prime}, n^{\prime}]$.  Use the definition of DFT, evaluating the output at the scaled coordinates bring in an extra $\alpha$ factor in the exponent term:
\begin{equation}
\begin{aligned}
& \mathrm{U} \left[m^{\prime}, n^{\prime}\right] 
\triangleq
\mathcal{F}\left\{\mathcal{P}_{\mathcal{F}}\left(\alpha x_c, \alpha y_c, 0, \omega \right)\right\}    \\
&= \sum_{m=-N / 2}^{N / 2-1} \sum_{n=-N / 2}^{N / 2-1} \mathrm{u}[m, n] \cdot \exp \left[\left(-j \frac{2 \pi}{N}\right) \cdot-\alpha\left(m \cdot m^{\prime}+n \cdot n^{\prime}\right)\right].
\end{aligned}
\label{eq:RSD_scale}
\end{equation}

However DFT can be slow to calculate, with some numerical procedure, we next derive how to rewrite the aforementioned equation in FFT. We start from the 1D case and then extend into 2D later.
\begin{equation}
\begin{aligned}
\mathrm{U}[n^{\prime}] 
& =\sum_{n=-N / 2}^{N / 2-1}\mathrm{u}[n] \cdot k_N^{-\alpha \cdot n \cdot n^{\prime}} \quad \text { where } \quad k_N=\exp \left[-j \frac{2 \pi}{N}\right]
\end{aligned}
\end{equation}
since $\quad \alpha \cdot n \cdot n^{\prime}=\alpha/2 \cdot \left( n ^2+ n ^{\prime 2}-\left(n ^{\prime}-n \right)^2 \right)$, Eq. \ref{eq:RSD_scale} can be rewritten as:
\begin{equation}
\begin{aligned}
\mathrm{U}[n^{\prime}] 
&= \sum_{n=-N / 2}^{N / 2-1} \mathrm{u}[n] \cdot k_N^{\frac{\alpha}{2}\left(n ^{\prime}-n \right)^2} \cdot k_N^{-\frac{\alpha}{2} n ^2} \cdot k_N^{-\frac{\alpha}{2} n ^{\prime 2}} \\
& =k_N^{-\frac{\alpha}{2} n^{\prime 2}} \sum_{n=-N / 2}^{N / 2-1} \underbrace{\mathrm{u}[n] \cdot k_N^{-\frac{\alpha}{2} n ^2}}_{\mathrm{u}^{\prime}[n]} \cdot \underbrace{k_N^{\frac{\alpha}{2}\left(n ^{\prime}-n \right)^2}}_{k_N^{\prime}\left[n ^{\prime}-n \right]} \\
& =k_N^{-\frac{\alpha}{2} n^{\prime 2}} \sum_{n =-N / 2}^{N / 2-1} \mathrm{u}^{\prime}[n] \cdot k_N^{\prime}\left[n ^{\prime}-n \right] 
\end{aligned}
\label{eq:RSD_scaleconv}
\end{equation}

After defining $\mathrm{u}^{\prime}[n] = \mathrm{u}[n]\cdot k_N^{-\frac{\alpha}{2} n ^2} $ and  $ k_N^{\prime}[n] = k_N^{\frac{\alpha}{2} n^2} $, Eq. \ref{eq:RSD_scaleconv} can be treated as a convolution and apply FFT to calculate.
\begin{equation}
\begin{aligned}
\mathrm{U}[n^{\prime}] 
&= k_N^{-\frac{\alpha}{2} n^{\prime 2}} \cdot 
\operatorname{IFFT}\left\{\operatorname{FFT}\left\{\mathrm{u}^{\prime}[n]\right\} \cdot \operatorname{FFT}\left\{k_N^{\prime}[n]\right\}\right\}
\end{aligned}
\end{equation}

\subsection{Volume of Frustum}
\label{sec:Volume_Frustum}
For a pyramidal frustum, the side length varies linearly with $z$ so that:

\begin{equation}
\begin{aligned}
x_{out} = x_{in} + \frac{z}{\alpha}  \\
y_{out} = y_{in} + \frac{z}{\beta} \\
z_{out} = z_{in} + z \\
\end{aligned}
\end{equation}

Then we can derive the volume by integrating cross-sectional areas along z:

\begin{equation}
\begin{aligned}
V_F &= \int_{z_{in}}^{z_{out}} A(z) dz \\
&= \int_{z_{in}}^{z_{out}} \left(x_{in} + \frac{z}{\alpha}\right) \left(y_{in} + \frac{z}{\beta}\right) dz \\
&= \int_{z_{in}}^{z_{out}}\left( x_{in}y_{in} + x_{in}\frac{z}{\beta} + y_{in}\frac{z}{\alpha}  +  \left(\frac{z^2}{\alpha\beta}\right) \right) dz \\
&= \left|  \left(z  x_{in}y_{in} + \frac{z^2}{2} \left(\frac{x_{in}}{\alpha} + \frac{y_{in}}{\beta} \right)  + \left(\frac{z^3}{3\alpha \beta }\right) \right) \right|_{z_{in}}^{z_{out}} \\
\end{aligned}
\end{equation}

Assuming that each cross-section of the frustum is a square, then $x_{in} = y_{in}$ and $\alpha = \beta$, we get:

\begin{equation}
\begin{aligned}
V_F &= \left|  \left(z x_{in}^2 + x_{in}\frac{z^2}{\alpha} + \left(\frac{z^3}{3\alpha^2}\right) \right) \right|_{z_{in}}^{z_{out}} \\
\end{aligned}
\end{equation}

\subsection{Sampling Proof}
\label{sec:NLOS_CS_Proof}
Suppose we have a phasor field point source in the hidden scene. The purpose of this section is to demonstrate that the contribution of this point source changes slowly along the transverse dimension on the relay wall. This means we can sample at a rate that is lower than that of the Nyquist Criterion without substantial loss of information.

Let's define the Nyquist Criterion. Assume you have a function $g(x)$ that is band-limited in the frequency domain s.t. 

\begin{align}
\mathcal{F} \{{g(x)}\} = G(f) \qquad G(f) = 0, \text{ if } f \geq |B|
\end{align}

where the $\mathcal{F}$ denotes the fourier transform operator. Then, the Nyquist criteria states that we can reconstruct the sampled signal from its spectrum provided we fix the sampling frequency to:

\begin{align}
    f_s \geq 2B  \qquad \rightarrow \qquad x_s \leq \frac{1}{2B}  
\end{align}

Let's operate in flatland with a point source located at $(x_0, z_0)$, and let's assume we have a relay wall at governed by the coordinates $(x_r, z_r)$. This simplifies the analysis, and is easy to extend to 3D. Let's assume that our point source radiates spherical waves \cite{reza_phasor_2019-1}, and we can calculate the contribution of a point source at $(x, z)$ using:

\begin{align}
P(r) = e^{ikr}
\end{align}

where

\begin{align}
r = \sqrt{(x - x_0)^2 + (z-z_0)^2}
\end{align}

and k, the wavenumber, is a spatial frequency in units of radians i.e. the rotations per wavelength:

\begin{align}
     k  = \frac{2 \pi}{\lambda}
\end{align}

In the phasor field formulation, each point in the hidden scene emits a gaussian pulse in time. In the frequency domain, we can write this as a sum of weighted point sources radiating at different frequencies:

\begin{align}
P(k) = \sum_k \alpha(k)e^{ikr}
\end{align}

Let $k^*$ define some maximum frequency of the gaussian pulse. Then, clearly our function is bandlimited and:

\begin{align}
P(k) = 0, \text{ if } k \geq |k^*|
\end{align}

And we can always recover the function without any loss of information provided:

\begin{align}
    k_s \geq 2k^*  \qquad \rightarrow \qquad x_s \leq \frac{1}{2k^*}  
\end{align}

This gives us an upper bound in terms of the sampling rate in units of radians i.e. rotations per unit second. We can convert the same criterion to spatial units and derive a nyquist criterion in terms of the wavelength, $\lambda^* = 2\pi/k^*$:

\begin{equation}
\begin{aligned}
     \lambda_s &= 2 \pi x_s \\
               & \leq 2 \pi \frac{1}{2k^*} = \frac{\lambda^*}{2} 
\end{aligned}
\end{equation}

where $\lambda_s$ is the minimum sampling interval. We can simplify the radial term:

\begin{equation}
\begin{aligned}
r &= \sqrt{(x - x_0)^2 + (z-z_0)^2} \\
  &= (z-z_0) \sqrt{ 1 + \frac{(x - x_0)^2}{(z-z_0)^2}} \\
  &\approx (z-z_0) + \frac{(x - x_0)^2}{2(z-z_0)}
\end{aligned}
\end{equation}

where we have made the paraxial approximation since it has generated reasonable results, especially a few meters away from the relay surface \cite{liu_non-line--sight_2019}. Now our kernel can be simplified as:

\begin{align}
P(x,z) &= e^{ik(z-z_0)}e^{ik\frac{(x - x_0)^2}{2(z-z_0)}}
\end{align}
We can parameterize our relay wall as $z = z_r$ plane from $-x_r$ to $x_r$. Now we want to calculate the contribution of this point source on the relay wall:

\begin{equation}
\begin{aligned}
P(x_r, z_r) &= e^{ik(z_r-z_0)}e^{ik\frac{(x_r - x_0)^2}{2(z_r-z_0)}} \\
& = e^{ik(z_r-z_0)}e^{ik''(x_r - x_0)} 
\end{aligned}
\end{equation}

where $k'' = \frac{k(x_r - x_0)}{2(z_r-z_0)}$. We can see that in the Fresnel regime, the phase changes linearly with changing the depth of the point source relative to the relay wall i.e. $(z_r-z_0)$. This means we have to sample at the nyquist rate or higher along the depth i.e. $\lambda_{sz} = \lambda^*/2$. 

But what about the transverse dimension? We can define a similar criterion as before for the transverse sampling interval $\lambda_{sx}$:

\begin{equation}
\begin{aligned}
     \lambda_{sx} & \leq 2 \pi \frac{1}{2} \frac{1}{k''}   \\
               & =  2 \pi \frac{1}{2} \frac{2(z_r - z_0) }{(x_r - x_0)k} \\
               & = \lambda^* \frac{(z_r - z_0) }{(x_r - x_0)} 
\end{aligned}
\end{equation}

Let the transverse sampling interval, $\lambda_{sx}$, be larger than the one along $z$, $\lambda_{sz}$. Imposing $\lambda_{sx} > \lambda_{sz}$ under the Fresnel approximation leads to: 
\begin{align}
\frac{(z_r - z_0) }{(x_r - x_0)}  & > \frac{1}{2}
\end{align}

In fact, we can impose a downsampling rate D as the ratio between the sampling intervals in $x$ relative to $z$ and figure out the regimes  $(|x_r - x_0|, |z_r-z_0|)$ for which this constraint is valid. 

\begin{align}
\frac{\lambda_{sx}}{\lambda_{sz}} =  \frac{2|z_r - z_0|}{|x_r-x_0|} > D  
\label{eq:upsample}
\end{align}

We note that $D$ can be any real number since it is a ratio between sampling intervals, and must be larger than 1 to compress the measurement in the transverse dimension. Small lateral offsets, \(|x_r-x_0| <  1\), allow for greater degree of downsampling. In 3D, a similar analysis can be used to find different downsampling constraints $D_x$ and $D_y$ for $x$ and $y$ respectively.

\subsubsection{Confocal Systems} Our previous analysis applies to non-confocal setups, where the illumination and detection locations are independent. In this case, the phase contribution from the relay wall to the point source can be ignored. In confocal setups, however, the hidden scene is both illuminated and detected from the same position on the relay surface. As a result, any changes in the relay surface's location affect not only the phase contribution from the relay wall to the hidden point but also the phase contribution from the hidden point back to the detection location on the surface. Consequently, the phasor field point source radiates as:

\begin{align}
P(r) = e^{ik2r}
\end{align}

Following the previous analysis, we get:

\begin{align}
P(x_r, z_r) &= e^{i2k(z_r-z_0)}e^{i2k\frac{(x_r - x_0)^2}{2(z_r-z_0)}}
\end{align}

Defining $k_{\text{con}} = 2k$, we get the same downsampling constraint as in Eq \ref{eq:upsample}. The only difference is that the baseline sampling rate for confocal setups is double that of non-confocal. 

\section{Hardware}
\label{sec:Hardware}

\begin{figure}
    \centering
    \captionsetup{skip=3pt}
    \includegraphics[width=0.7\columnwidth]{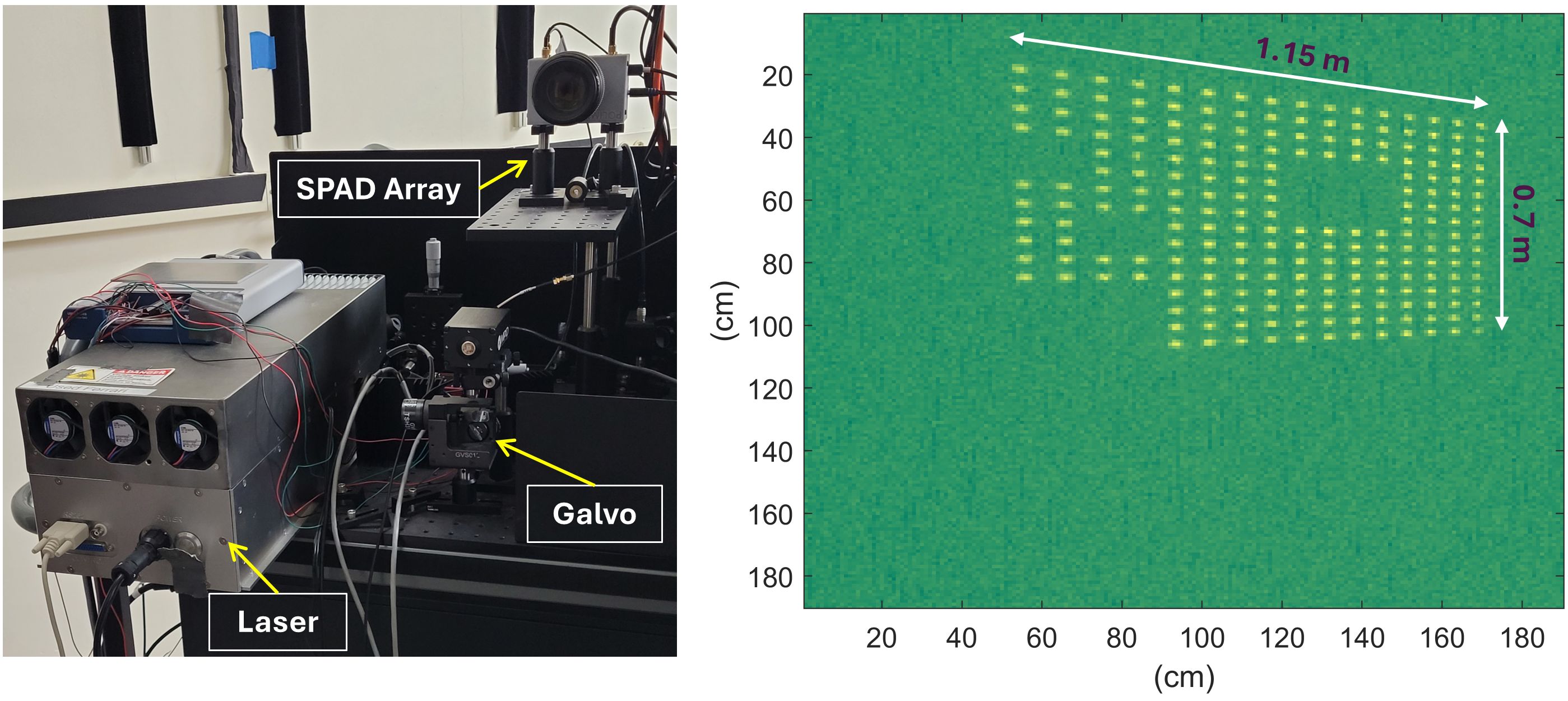}
    \caption{An image of our NLOS imaging system consisting of a pulsed laser, galvonometer, and SPAD array on the left. The image on the right displays how focusing the SPAD array to a large area (1.15 m x 0.7 m) on the planar 1.9 m x 1.9 m relay surface generates a non-uniform grid. The empty spaces correspond to 42 hot pixels in the array.}
    \label{fig:Hardware}
\end{figure}

In this work, we build an active NLOS imaging system with a \text{PM-1.03-25\textsuperscript{TM}} pulsed laser from Polar Laser Laboratories laser as the illumination source, and a customized gated 16 x 16 pixel Single Photon Avalanche Diode (SPAD) array, developed at Politecnico di Milano \cite{riccardo_fast-gated_2022}, as the detection source (Column 1, Fig. \ref{fig:Hardware}).  The laser is coupled with a frequency doubler to generate 515 nm pulses at an average rep rate of 5 MHz and average power of 375 mW. We sequentially illuminate individual locations on a 1.9 m x 1.9 m area on a relay wall using a two-mirror Thorlabs galvanometer (Thorlabs GVS012). Furthermore, we use an Edmund Optics 6X Manual Zoom Video lens to focus our 16x16 array to a 50 cm x 32.5 cm (Width x Height) area on the relay wall. For the results demonstrated in Fig. \ref{fig:NURSD1_SPADGRID_SUPP}, we use a  Fujinon 3 MP Varifocal Lens (3.8-13 mm, 3.4x Zoom) to focus our 16x16 array to a 1.1 m x 0.7 m area on the relay wall (Column 2, Fig. \ref{fig:Hardware}). The overall temporal resolution of our system is characterized by a Full-Width at Half Maximum (FWHM) of approximately 65 ps.

\section{Supplemental Results}
\label{sec:Supp_Results}

\subsection{Detector Arrays}
\label{sec:NURSD_SPADArray}
\begin{figure}
    \centering
    \captionsetup{skip=3pt}
    \includegraphics[width=0.7\columnwidth]{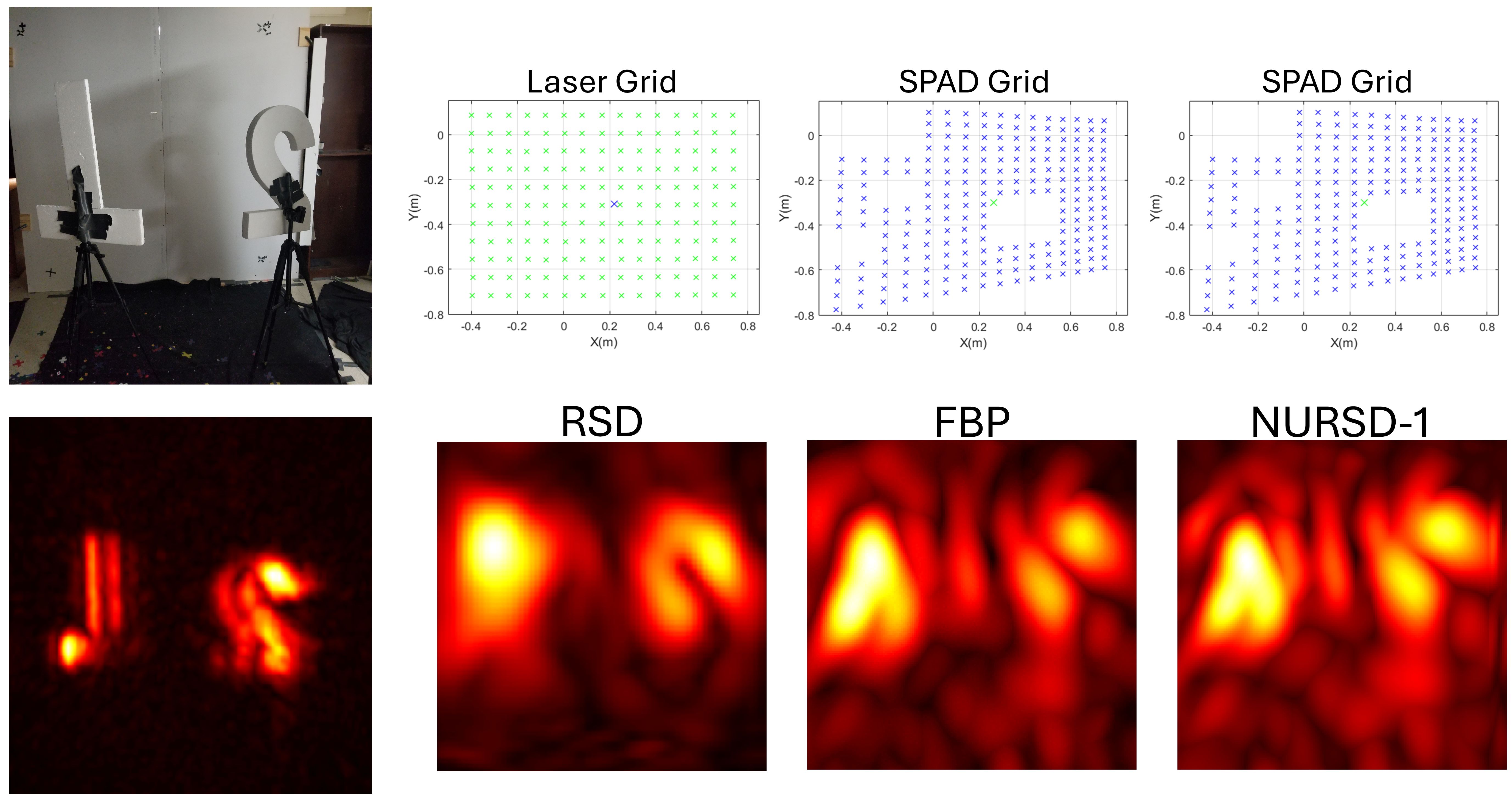}
    \caption{Column 1 shows an image of the hidden scene and the Standard RSD reconstruction when the full 1.9 m x 1.9 m laser grid is used to acquire the dataset. Column 2 shows the standard RSD reconstruction when using a subset of the laser grid that is approximately equal to the SPAD grid. Columns 3 and 4 show the FBP and NURSD reconstructions, respectively, for this SPAD grid. The acquisition time for the SPAD grid is ~250 times less than the equivalent laser grid due to parallel acquisition}
\label{fig:NURSD1_SPADGRID_SUPP}
\end{figure}

While sensing arrays enable parallel acquisition, achieving uniform spacing on the relay surface can be challenging, especially when focusing from an angle and due to hot pixels causing additional non-uniformities. In this section, we focus a 16x16 SPAD array (with 216 active pixels) over a large area and collect a dataset where the hidden scene consists of the letters "2" and "T". We then reconstruct the scene using both FBP and NURSD-1 for a single illumination position, showing that the two reconstructions are consistent with each other (Columns 3 and 4 of Fig \ref{fig:NURSD1_SPADGRID_SUPP}). This SPAD grid is a subset of the full laser grid (Column 2, Fig. \ref{fig:Hardware}), which is why the bottom of the 2 and T are missing relative to Column 1, which is reconstructed using the full laser grid. Column 2 generates a reconstruction using the Standard RSD for a subset of the laser grid roughly equivalent to the SPAD grid and a single detection position. We note that the laser grid in Column 2 takes 250x longer to acquire than the equivalent SPAD grid in Columns 3 and 4. Since the SPAD grid was collected with a 1ms exposure time, we need to use a large wavelength to clearly discern the hidden objects.

\subsection{4D Light Transport}
\label{sec:4D_LightTransport}

The Standard RSD algorithm can be used to generate a 4D video of our virtual source propagating and interacting with the hidden scene \cite{liu_phasor_2020}. This algorithm can be adapted to extract complex light transport of the hidden scene \cite{sultan2024iteratingtransientlighttransport}. 

\begin{equation}
\begin{aligned}
I\left(\vec{x}_{v}, t\right)
& =\iiint \mathcal{P_F}(\vec{x}_{p}, \omega) e^{-j \frac{\omega}{c}\left(\left|\vec{x}_{c}-\vec{x}_{v}\right|\right)} e^{j \omega t} d\omega  d \vec{x}_{c}  d \vec{x}_{p} 
\label{eq:Video_Voxel_Freq_Offset}
\end{aligned}
\end{equation}

\begin{figure}
    \centering
    \captionsetup{skip=3pt}
\includegraphics[width=1\columnwidth]{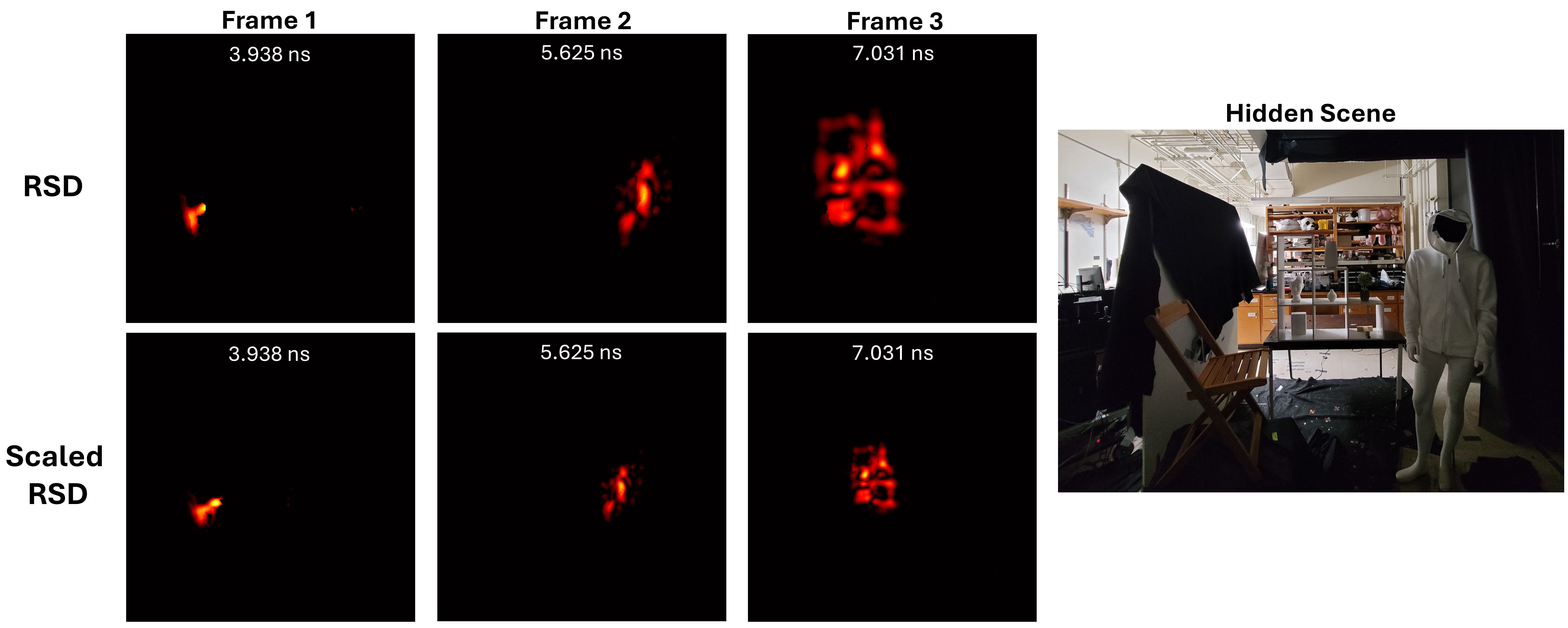}
    \caption{We generate 4D videos of light transport in the hidden scene using the RSD (Row 1) and the Scaled RSD (Row 2).The mannequin and shelf towards the back of the hidden volume appear smaller in the scaled RSD (Row 2, Columns 3 and 4) due to perspective projection}
    \label{fig:SRSD_Video}
\end{figure}

In Row 2 of Fig \ref{fig:SRSD_Video}, we show that the Scaled RSD can be used to generate these 4D videos of the hidden scene, while maintaining perspective projection.

\subsection{Fusions}
\label{sec:Fusions}

\begin{figure*}
    \centering
    \captionsetup{skip=3pt}
    \includegraphics[width=1\textwidth]{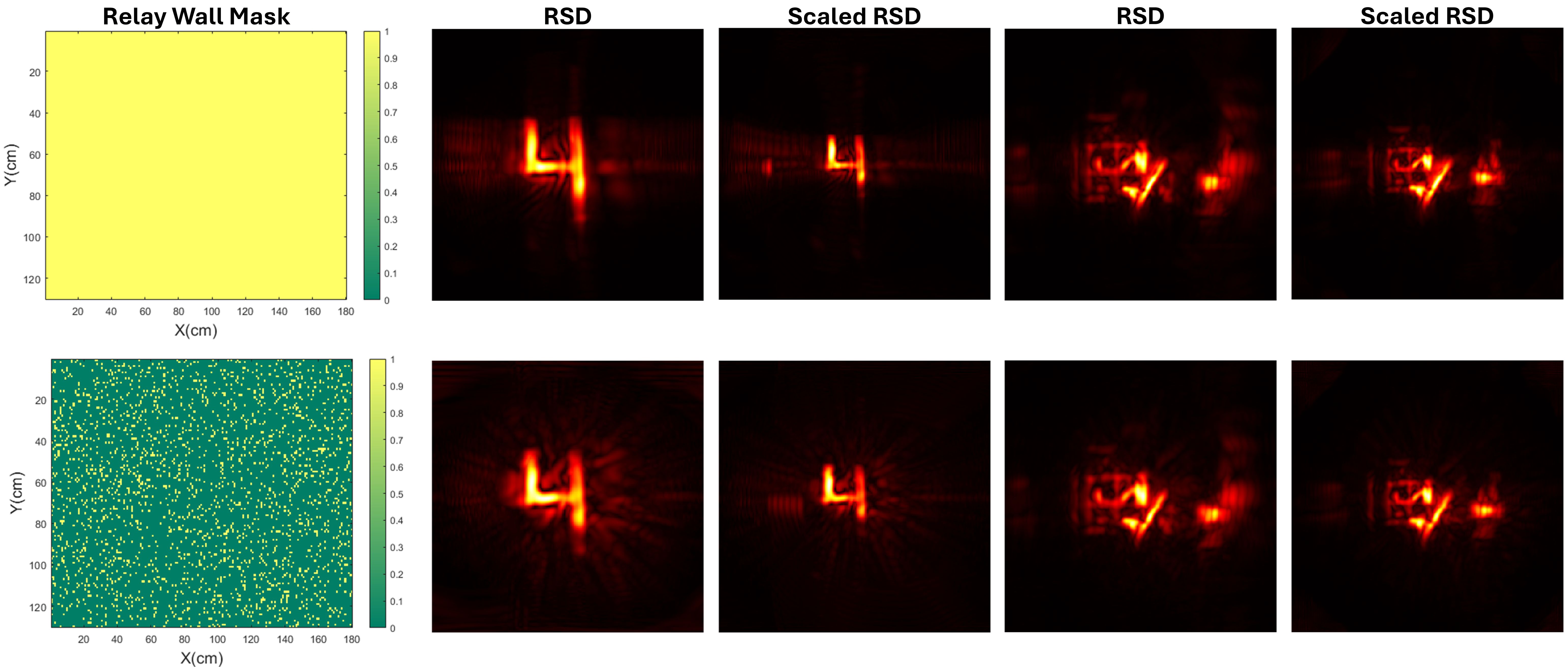}
    \caption{In Row 1, we show the reconstructions for two different datasets using the Standard RSD (Columns 2 and 4) and Scaled RSD (Columns 3 and 5). In Row 2, we randomly subsample 10$\%$ of the relay surface (Column 1) for the same datasets and interpolate to a uniform grid. We show that both Standard RSD (Columns 2 and 4) and the Scaled RSD (Columns 3 and 5) reconstructions do not degrade in quality after discarding 90$\%$ of grid positions. Scaled RSD generates reconstructions with perspective projection.}
    \label{fig:SRSD_NonUniform}
\end{figure*}

We can combine existing algorithms to generate novel ones: 
\paragraph{SRSD + NURSD-2}  

\begin{equation}
 \begin{split} 
\mathcal{P}_{\mathcal{F}} \left( x_\ell, y_\ell, z_v, \omega \right) 
&= \\
\operatorname{NUFFT-2}\Bigl\{ \operatorname{SFFT}\left\{\mathcal{P}_{\mathcal{F}} \left( m \Delta x, n\Delta y, 0, \omega \right) \right\}  & \cdot  \operatorname{FFT}\left\{G\left(\alpha m  \Delta x, \beta n\Delta y, z_v, \omega\right) \right\}\Bigr\}
\label{eq:NURSD2_SRSD}
\end{split}   
\end{equation}

where SFFT is the scaled Fourier transform introduced in Eq \ref{eq:SFFT}. This enables us to generate reconstructions where we sample the scaled voxel grid with non-uniform spacing.

\paragraph{3D RSD/3D NURSD + SRSD} The second stage is simply the standard RSD algorithm, so it can be replaced by SRSD to generate reconstructions where the voxel size increases at a constant rate at each depth.

\paragraph{3D RSD/3D NURSD + NURSD-2}
The second stage is simply the standard RSD algorithm, so it can be replaced by NU-RSD 2 to generate reconstructions for arbitrary sampling schemes for the voxel grid.

\paragraph{SRSD on Non-Uniform Acquisition schemes}
The SRSD algorithm requires the input to be uniformly spaced. One strategy, that has worked for the 3D RSD, is to oversample the relay surface with a non-uniform grid, indicated by ${\vec{x}_{\mathrm{c}}}^\prime$, and interpolate to a regular grid $({\vec{x}_{\mathrm{c}}})$:

\begin{equation}
\begin{aligned}
& \mathcal{P_F} \left(\vec{x}_{\mathrm{c}}, \omega \right)
= \mathcal{\phi} \left( \mathcal{P_F} \left( {\vec{x}_{\mathrm{c}}}^\prime, \omega  \right) \right)
\end{aligned}
\end{equation}

Then, we can proceed with the SRSD as usual. In Fig. $\ref{fig:SRSD_NonUniform}$, we demonstrate that this strategy works well on experimental data. 

How is this different from the NUFFT-1? The NUFFT-1 can be written as:

\begin{equation}
\begin{aligned}
 \mathcal{P_F} \left(\vec{x}_{\mathrm{c}}, \omega \right) &= \operatorname{NUFFT-1}\left\{ \mathcal{P} \left( {\vec{x}_{\mathrm{c}}}^\prime, \omega \right) \right\}  \\ &= \mathcal{D} \left(  \operatorname{FFT} \left\{ \mathcal{\phi} \left(  \mathcal{B} \Bigl(  \mathcal{P} \left( {\vec{x}_{\mathrm{c}}}^\prime, \omega \right) \Bigr)  \right)  \right\} \right) 
\end{aligned}
\label{eq:3d_nursd_Int}
\end{equation}

where NUFFT-1 optimizes the interpolation for the FFT operator by incorporating blurring, $\mathcal{B}$, and deblurring, $\mathcal{D}$, operators for some blurring kernel, $\psi(x)$. Restating SFFT for $\alpha = \beta$, $m = n$ for simplicity:

\begin{equation}
\begin{aligned} 
 \mathcal{P^\alpha_F} \left(\vec{x}_{\mathrm{c}}, \omega \right) &= \operatorname{SFFT} \left\{ \mathcal{P} \left( {\vec{x}_{\mathrm{c}}}, \omega \right), \alpha \right\} \\
&=  k_M^{-\alpha m^{\prime 2}}
\operatorname{IFFT}\Bigl\{ \operatorname{FFT} \left\{\mathcal{P}_{\mathcal{F}} \left( \vec{x}_{\mathrm{c}}, \omega \right) k_M^{-\alpha m^{2}} \right\}  \cdot  \operatorname{FFT}\left\{k_M^{\alpha m^2}  \right\}\Bigr\}
\end{aligned}
\end{equation}

In principle, one should be able to fuse the two algorithms:

\begin{equation}
\begin{aligned} 
 \mathcal{P^\alpha_F}  \left(\vec{x}_{\mathrm{c}}^\prime, \omega \right) &= \operatorname{NU-SFFT} \left\{ \mathcal{P} \left( {\vec{x}_{\mathrm{c}}}^\prime, \omega \right), \alpha \right\} \\
& =  k_M^{-\alpha m^{\prime 2}}
\operatorname{IFFT}\Bigl\{ \mathcal{D} \left(  \operatorname{FFT} \left\{  \mathcal{\phi} \left[  \mathcal{B} \Bigl(  \mathcal{P} \left( {\vec{x}_{\mathrm{c}}}^\prime, \omega \right) \Bigr)  \right] k_M^{-\alpha m^{2}} \right\} \right)   \cdot  \operatorname{FFT}\left\{k_M^{\alpha m^2}  \right\}\Bigr\}
\end{aligned}
\end{equation}

However, this requires optimizing blurring kernel and interpolation method for the scaled grid, in addition to adjusting the existing implementation of NUFFT to incorporate a phase factor given by $k_M^{-\alpha m^{2}}$. This can be the subject of future work.

\subsection{NLOS Depth Resolution}
\begin{figure*}
    \centering
    \captionsetup{skip=3pt}
    \includegraphics[width=0.6\textwidth]{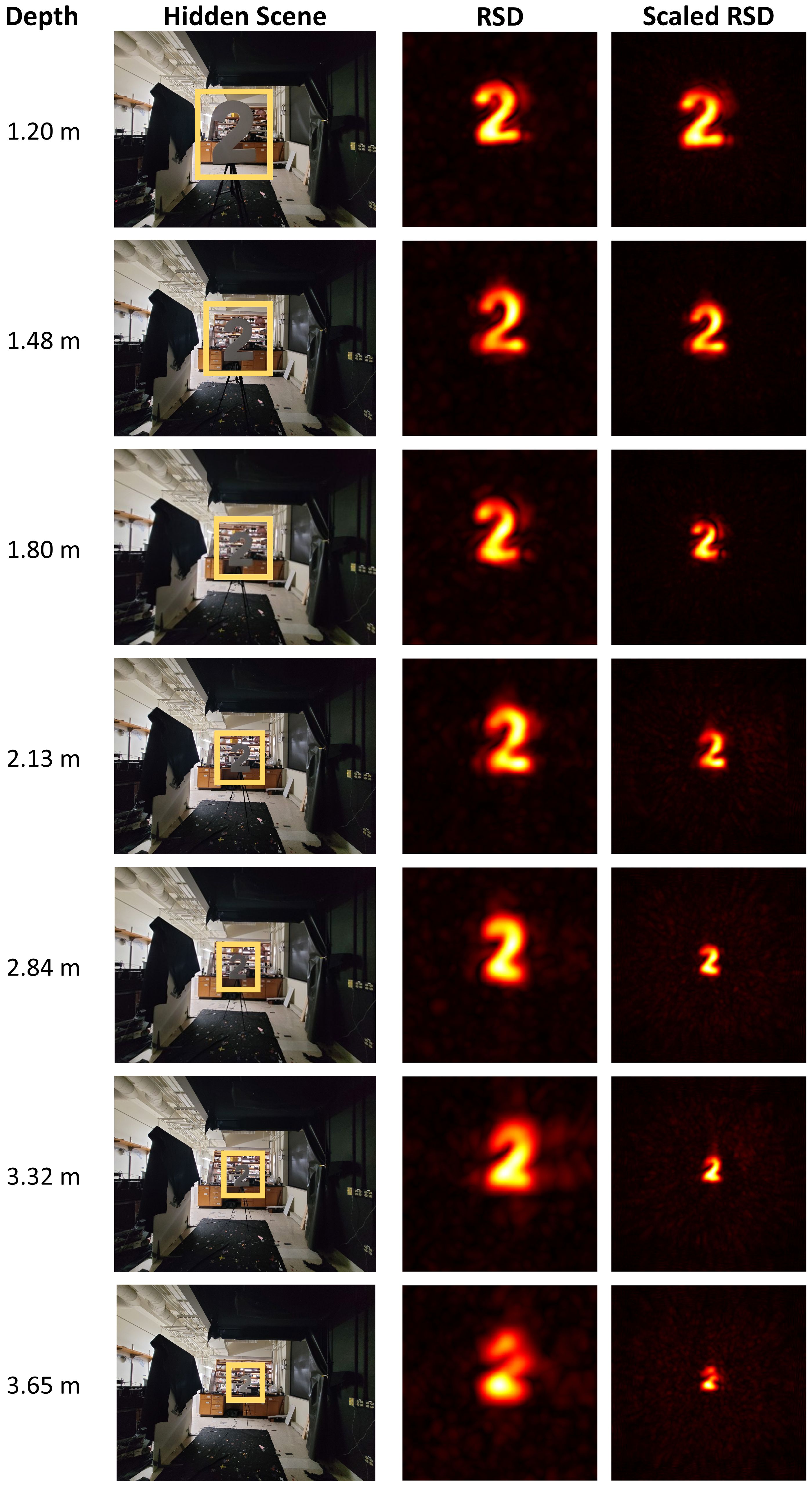}
    \caption{We place the digit "2" at various depths within the hidden scene. Column 1 indicates the average depth of the "2" for each row. Column 2 shows the smartphone image of the hidden scene, with a yellow box marking the location of the "2." Column 3 presents the Standard RSD reconstruction, while Column 4 features the Scaled RSD reconstruction. Both the reference image (Column 2) and the Scaled RSD reconstruction (Column 4) exhibit perspective projection, where the "2" appears smaller as its distance from the relay surface increases.}
    \label{fig:SRSD_Depth}
\end{figure*}

In this section, we demonstrate how resolution diminishes as the distance or depth from the relay surface increases. We collected NLOS datasets for the digit "2" placed at various depths within the hidden scene. Fig. \ref{fig:SRSD_Depth} summarizes the results: Column 1 displays the average depth of the hidden object, while Column 2 provides the reference image captured by a smartphone camera from the relay surface's perspective. Columns 3 and 4 present the reconstructed scenes using the Standard RSD and the Scaled RSD, respectively. The Standard RSD shows how resolution degrades as the digit "2" moves farther from the relay surface. In contrast, the Scaled RSD compensates for this loss by increasing the voxel size at greater depths, effectively shrinking the reconstructed object. This adjustment creates a perspective projection, replicating the effect observed in the smartphone reference image.

\end{document}